\documentclass[iop]{emulateapj}
\pdfoutput=1

\usepackage{natbib}
\usepackage{amsmath}
\usepackage{xcolor}

\newcommand{\bvec}[1]{\ensuremath{\boldsymbol{#1}}}

\newcommand{\CoGRA}{\alpha_{J2000} = 17^h48^m4.85^s\pm0.5^s}
\newcommand{\CoGDec}{\delta_{J2000} = -24^\circ46^m44.6^s\pm0.5^s}
\newcommand{\CoGRATuc}{\alpha_{J2000} = 00^h24^m05.36^s}
\newcommand{\CoGDECTuc}{\delta_{J2000} = -72^\circ04^m53.2^s}
\newcommand{\CoGl}{l = 3.8^\circ}
\newcommand{\CoGb}{b = 1.7^\circ}
\newcommand{\lanzonirc}{0.26\,{\rm pc}}
\newcommand{\mioccchirc}{0.22\pm0.01\,{\rm pc}}

\newcommand{\ag}{a_g = 5.1 \times 10^{-10} \pm 1.4 \times 10^{-10} \, {\rm m}\,{\rm s}^{-2}}
\newcommand{\q}{q=1.57\pm0.19}
\newcommand{\ntemp}{16}
\newcommand{\nwalkeraccel}{256}

\newcommand{\ndimaccel}{37}
\newcommand{\ndimjerk}{69}
\newcommand{\ndimbh}{70}

\newcommand{\NSmass}{M_X=1.33 \pm 0.07\,{\rm M}_\odot}
\newcommand{\qheinke}{q=1.43 \pm 0.11}
\newcommand{\tucrho}{\rho_c=(1.2_{-0.3}^{+0.3})\times10^5\,{\rm M}_\odot\,{\rm pc}^{-3}}
\newcommand{\tucrc}{r_c=0.66_{-0.08}^{+0.18}\,{\rm pc}}

\newcommand{\mbhupper}{M_{\rm BH}\simeq3\times10^4 \,{\rm M}_\odot}
\newcommand{\mbhcand}{M_{\rm BH}\simeq500 \,{\rm M}_\odot}
\newcommand{\Npsrtot}{37}
\newcommand{\Npsr}{36}

\newcommand{\denctuc}{(1.19_{-0.36}^{+0.61})}
\newcommand{\rctuc}{0.50_{-0.10}^{+0.14}}
\newcommand{\alphatuc}{-5.73_{-1.91}^{+1.57}}

\newcommand{\dencinit}{1.67_{-0.29}^{+0.37}}
\newcommand{\dencaccel}{1.44_{-0.21}^{+0.22}}
\newcommand{\dencjerk}{1.58_{-0.13}^{+0.13}}
\newcommand{\dencBH}{1.62_{-0.16}^{+0.18}}
\newcommand{\dencCOG}{1.41_{-0.14}^{+0.18}}
\newcommand{\rcinitarcsec}{5.94}
\newcommand{\rcinit}{0.17_{-0.02}^{+0.02}}
\newcommand{\rcaccel}{0.16_{-0.01}^{+0.01}}
\newcommand{\rcjerk}{0.16_{-0.01}^{+0.01}}
\newcommand{\rcBH}{0.15_{-0.01}^{+0.02}}
\newcommand{\rcCOG}{0.16_{-0.01}^{+0.01}}
\newcommand{\alphainit}{-3.71 \pm 0.57}
\newcommand{\alphaaccel}{-3.57_{-0.41}^{+0.45}}
\newcommand{\alphajerk}{-3.14_{-0.53}^{+0.52}}
\newcommand{\alphaBH}{-3.59_{-0.31}^{+0.33}}
\newcommand{\alphaCOG}{-3.56_{-0.56}^{+0.47}}
\newcommand{\qaccel}{1.52_{-0.14}^{+0.15}}

\newcommand{\zinitI}{\@m N/A}
\newcommand{\zprimaryaccelI}{1.1\times10^{-2} \pm 1.7\times10^{-3}}
\newcommand{\zsecondaryaccelI}{1.8\times10^{0} \pm 3.6\times10^{-2}}
\newcommand{\zprobaccelI}{230.0}
\newcommand{\zprimaryjerkI}{1.0\times10^{-2} \pm 1.2\times10^{-3}}
\newcommand{\zsecondaryjerkI}{1.6\times10^{0} \pm 1.5\times10^{-1}}
\newcommand{\zprobjerkI}{9.7}

\newcommand{\zinitae}{\@m N/A}
\newcommand{\zprimaryaccelae}{2.8\times10^{-1} \pm 5.3\times10^{-2}}
\newcommand{\zsecondaryaccelae}{1.0\times10^{-1} \pm 2.9\times10^{-2}}
\newcommand{\zprobaccelae}{1.3}
\newcommand{\zprimaryjerkae}{3.2\times10^{-1} \pm 2.7\times10^{-1}}
\newcommand{\zsecondaryjerkae}{8.6\times10^{-2} \pm 2.1\times10^{-2}}
\newcommand{\zprobjerkae}{1.3}

\newcommand{\zinitZ}{\@m N/A}
\newcommand{\zprimaryaccelZ}{1.5\times10^{-2} \pm 4.6\times10^{-3}}
\newcommand{\zsecondaryaccelZ}{1.1\times10^{0} \pm 1.2\times10^{-1}}
\newcommand{\zprobaccelZ}{230.0}
\newcommand{\zprimaryjerkZ}{1.4\times10^{-2} \pm 3.7\times10^{-3}}
\newcommand{\zsecondaryjerkZ}{7.4\times10^{-1} \pm 1.8\times10^{-1}}
\newcommand{\zprobjerkZ}{14.0}

\newcommand{\zinitW}{\@m N/A}
\newcommand{\zprimaryaccelW}{-8.6\times10^{-3} \pm 1.3\times10^{-3}}
\newcommand{\zsecondaryaccelW}{-5.4\times10^{-2} \pm 4.3\times10^{-4}}
\newcommand{\zprobaccelW}{230.0}
\newcommand{\zprimaryjerkW}{-7.8\times10^{-3} \pm 9.5\times10^{-4}}
\newcommand{\zsecondaryjerkW}{-1.7\times10^{0} \pm 2.3\times10^{-1}}
\newcommand{\zprobjerkW}{4.9}

\newcommand{\zinitab}{\@m N/A}
\newcommand{\zprimaryaccelab}{-3.2\times10^{-2} \pm 1.3\times10^{-1}}
\newcommand{\zsecondaryaccelab}{-3.2\times10^{-2} \pm 1.3\times10^{-1}}
\newcommand{\zprobaccelab}{1.0}
\newcommand{\zprimaryjerkab}{4.0\times10^{-2} \pm 1.9\times10^{-1}}
\newcommand{\zsecondaryjerkab}{4.0\times10^{-2} \pm 1.9\times10^{-1}}
\newcommand{\zprobjerkab}{1.0}

\newcommand{\zinitY}{\@m N/A}
\newcommand{\zprimaryaccelY}{-3.0\times10^{-2} \pm 4.6\times10^{-3}}
\newcommand{\zsecondaryaccelY}{-8.7\times10^{-1} \pm 8.3\times10^{-2}}
\newcommand{\zprobaccelY}{2.1}
\newcommand{\zprimaryjerkY}{-2.7\times10^{-2} \pm 5.2\times10^{-3}}
\newcommand{\zsecondaryjerkY}{-8.0\times10^{-1} \pm 1.1\times10^{-1}}
\newcommand{\zprobjerkY}{1.0}

\newcommand{\zinitP}{\@m N/A}
\newcommand{\zprimaryaccelP}{-6.5\times10^{-2} \pm 3.6\times10^{-2}}
\newcommand{\zsecondaryaccelP}{-4.9\times10^{-1} \pm 1.7\times10^{-1}}
\newcommand{\zprobaccelP}{10.0}
\newcommand{\zprimaryjerkP}{7.4\times10^{-2} \pm 1.4\times10^{-1}}
\newcommand{\zsecondaryjerkP}{7.4\times10^{-2} \pm 1.4\times10^{-1}}
\newcommand{\zprobjerkP}{1.0}

\newcommand{\zinitM}{\@m N/A}
\newcommand{\zprimaryaccelM}{-1.0\times10^{-1} \pm 1.6\times10^{-2}}
\newcommand{\zsecondaryaccelM}{-4.1\times10^{-1} \pm 5.0\times10^{-2}}
\newcommand{\zprobaccelM}{3.0}
\newcommand{\zprimaryjerkM}{-8.8\times10^{-2} \pm 1.1\times10^{-2}}
\newcommand{\zsecondaryjerkM}{-4.3\times10^{-1} \pm 6.1\times10^{-2}}
\newcommand{\zprobjerkM}{3.6}

\newcommand{\zinitR}{\@m N/A}
\newcommand{\zprimaryaccelR}{-7.0\times10^{-2} \pm 1.3\times10^{-1}}
\newcommand{\zsecondaryaccelR}{-7.0\times10^{-2} \pm 1.3\times10^{-1}}
\newcommand{\zprobaccelR}{1.0}
\newcommand{\zprimaryjerkR}{5.5\times10^{-2} \pm 1.2\times10^{-1}}
\newcommand{\zsecondaryjerkR}{5.5\times10^{-2} \pm 1.2\times10^{-1}}
\newcommand{\zprobjerkR}{1.0}

\newcommand{\zinitO}{\@m N/A}
\newcommand{\zprimaryaccelO}{6.6\times10^{-2} \pm 8.7\times10^{-3}}
\newcommand{\zsecondaryaccelO}{8.2\times10^{-1} \pm 1.1\times10^{-1}}
\newcommand{\zprobaccelO}{23.0}
\newcommand{\zprimaryjerkO}{5.7\times10^{-2} \pm 8.3\times10^{-3}}
\newcommand{\zsecondaryjerkO}{7.4\times10^{-1} \pm 9.9\times10^{-2}}
\newcommand{\zprobjerkO}{8.0}

\newcommand{\zinitF}{\@m N/A}
\newcommand{\zprimaryaccelF}{1.4\times10^{-3} \pm 2.4\times10^{-4}}
\newcommand{\zsecondaryaccelF}{1.4\times10^{-3} \pm 2.4\times10^{-4}}
\newcommand{\zprobaccelF}{1.0}
\newcommand{\zprimaryjerkF}{1.4\times10^{-3} \pm 2.0\times10^{-4}}
\newcommand{\zsecondaryjerkF}{1.4\times10^{-3} \pm 2.0\times10^{-4}}
\newcommand{\zprobjerkF}{1.0}

\newcommand{\zinitah}{\@m N/A}
\newcommand{\zprimaryaccelah}{-1.2\times10^{-1} \pm 3.6\times10^{-2}}
\newcommand{\zsecondaryaccelah}{-4.6\times10^{-1} \pm 1.6\times10^{-1}}
\newcommand{\zprobaccelah}{8.2}
\newcommand{\zprimaryjerkah}{-1.1\times10^{-1} \pm 1.9\times10^{-1}}
\newcommand{\zsecondaryjerkah}{-1.1\times10^{-1} \pm 1.9\times10^{-1}}
\newcommand{\zprobjerkah}{1.0}

\newcommand{\zinitaf}{\@m N/A}
\newcommand{\zprimaryaccelaf}{1.1\times10^{-1} \pm 5.5\times10^{-2}}
\newcommand{\zsecondaryaccelaf}{5.4\times10^{-1} \pm 1.4\times10^{-1}}
\newcommand{\zprobaccelaf}{10.0}
\newcommand{\zprimaryjerkaf}{9.5\times10^{-2} \pm 3.6\times10^{-2}}
\newcommand{\zsecondaryjerkaf}{3.3\times10^{-1} \pm 1.4\times10^{-1}}
\newcommand{\zprobjerkaf}{1.1}

\newcommand{\zinitU}{\@m N/A}
\newcommand{\zprimaryaccelU}{-1.3\times10^{-1} \pm 3.0\times10^{-2}}
\newcommand{\zsecondaryaccelU}{-5.3\times10^{-1} \pm 8.6\times10^{-2}}
\newcommand{\zprobaccelU}{2.0}
\newcommand{\zprimaryjerkU}{-5.6\times10^{-1} \pm 1.5\times10^{-1}}
\newcommand{\zsecondaryjerkU}{-1.3\times10^{-1} \pm 4.1\times10^{-2}}
\newcommand{\zprobjerkU}{1.1}

\newcommand{\zinitL}{\@m N/A}
\newcommand{\zprimaryaccelL}{2.0\times10^{-2} \pm 2.8\times10^{-2}}
\newcommand{\zsecondaryaccelL}{2.0\times10^{-2} \pm 2.8\times10^{-2}}
\newcommand{\zprobaccelL}{1.0}
\newcommand{\zprimaryjerkL}{1.5\times10^{-2} \pm 8.5\times10^{-2}}
\newcommand{\zsecondaryjerkL}{2.7\times10^{0} \pm 2.7\times10^{-1}}
\newcommand{\zprobjerkL}{26.0}

\newcommand{\zinitN}{\@m N/A}
\newcommand{\zprimaryaccelN}{-7.8\times10^{-2} \pm 8.2\times10^{-3}}
\newcommand{\zsecondaryaccelN}{-9.1\times10^{-1} \pm 8.1\times10^{-2}}
\newcommand{\zprobaccelN}{8.8}
\newcommand{\zprimaryjerkN}{-7.2\times10^{-2} \pm 7.2\times10^{-3}}
\newcommand{\zsecondaryjerkN}{-9.6\times10^{-1} \pm 9.8\times10^{-2}}
\newcommand{\zprobjerkN}{12.0}

\newcommand{\zinitag}{\@m N/A}
\newcommand{\zprimaryaccelag}{-5.2\times10^{-4} \pm 7.1\times10^{-5}}
\newcommand{\zsecondaryaccelag}{-5.2\times10^{-4} \pm 7.1\times10^{-5}}
\newcommand{\zprobaccelag}{1.0}
\newcommand{\zprimaryjerkag}{-4.9\times10^{-4} \pm 8.2\times10^{-5}}
\newcommand{\zsecondaryjerkag}{-4.9\times10^{-4} \pm 8.2\times10^{-5}}
\newcommand{\zprobjerkag}{1.0}

\newcommand{\zinitaj}{\@m N/A}
\newcommand{\zprimaryaccelaj}{5.1\times10^{-2} \pm 2.2\times10^{-1}}
\newcommand{\zsecondaryaccelaj}{5.1\times10^{-2} \pm 2.2\times10^{-1}}
\newcommand{\zprobaccelaj}{1.0}
\newcommand{\zprimaryjerkaj}{5.3\times10^{-2} \pm 1.0\times10^{-1}}
\newcommand{\zsecondaryjerkaj}{5.3\times10^{-2} \pm 1.0\times10^{-1}}
\newcommand{\zprobjerkaj}{1.0}

\newcommand{\zinitV}{\@m N/A}
\newcommand{\zprimaryaccelV}{9.0\times10^{-2} \pm 1.7\times10^{-2}}
\newcommand{\zsecondaryaccelV}{8.9\times10^{-1} \pm 1.7\times10^{-1}}
\newcommand{\zprobaccelV}{11.0}
\newcommand{\zprimaryjerkV}{1.3\times10^{0} \pm 6.2\times10^{-1}}
\newcommand{\zsecondaryjerkV}{8.0\times10^{-2} \pm 3.2\times10^{-2}}
\newcommand{\zprobjerkV}{1.8}

\newcommand{\zinitC}{\@m N/A}
\newcommand{\zprimaryaccelC}{1.5\times10^{-1} \pm 2.2\times10^{-2}}
\newcommand{\zsecondaryaccelC}{6.1\times10^{-1} \pm 1.5\times10^{-1}}
\newcommand{\zprobaccelC}{1.8}
\newcommand{\zprimaryjerkC}{2.7\times10^{-1} \pm 1.8\times10^{-1}}
\newcommand{\zsecondaryjerkC}{2.7\times10^{-1} \pm 1.8\times10^{-1}}
\newcommand{\zprobjerkC}{1.0}

\newcommand{\zinitG}{\@m N/A}
\newcommand{\zprimaryaccelG}{2.8\times10^{-2} \pm 1.7\times10^{-1}}
\newcommand{\zsecondaryaccelG}{2.8\times10^{-2} \pm 1.7\times10^{-1}}
\newcommand{\zprobaccelG}{1.0}
\newcommand{\zprimaryjerkG}{2.2\times10^{-2} \pm 1.8\times10^{-2}}
\newcommand{\zsecondaryjerkG}{1.9\times10^{0} \pm 6.0\times10^{-1}}
\newcommand{\zprobjerkG}{1.1}

\newcommand{\zinitai}{\@m N/A}
\newcommand{\zprimaryaccelai}{-9.8\times10^{-2} \pm 1.7\times10^{-2}}
\newcommand{\zsecondaryaccelai}{-9.5\times10^{-1} \pm 2.4\times10^{-1}}
\newcommand{\zprobaccelai}{3.1}
\newcommand{\zprimaryjerkai}{-9.0\times10^{-2} \pm 2.4\times10^{-2}}
\newcommand{\zsecondaryjerkai}{-6.5\times10^{-1} \pm 3.0\times10^{-1}}
\newcommand{\zprobjerkai}{8.4}

\newcommand{\zinitK}{\@m N/A}
\newcommand{\zprimaryaccelK}{1.3\times10^{-1} \pm 2.5\times10^{-1}}
\newcommand{\zsecondaryaccelK}{1.3\times10^{-1} \pm 2.5\times10^{-1}}
\newcommand{\zprobaccelK}{1.0}
\newcommand{\zprimaryjerkK}{1.2\times10^{-1} \pm 2.1\times10^{-1}}
\newcommand{\zsecondaryjerkK}{1.2\times10^{-1} \pm 2.1\times10^{-1}}
\newcommand{\zprobjerkK}{1.0}

\newcommand{\zinitaa}{\@m N/A}
\newcommand{\zprimaryaccelaa}{3.5\times10^{-1} \pm 1.1\times10^{-1}}
\newcommand{\zsecondaryaccelaa}{3.5\times10^{-1} \pm 1.1\times10^{-1}}
\newcommand{\zprobaccelaa}{1.0}
\newcommand{\zprimaryjerkaa}{3.5\times10^{-1} \pm 1.2\times10^{-1}}
\newcommand{\zsecondaryjerkaa}{3.5\times10^{-1} \pm 1.2\times10^{-1}}
\newcommand{\zprobjerkaa}{1.0}

\newcommand{\zinitH}{\@m N/A}
\newcommand{\zprimaryaccelH}{6.2\times10^{-2} \pm 2.4\times10^{-1}}
\newcommand{\zsecondaryaccelH}{6.2\times10^{-2} \pm 2.4\times10^{-1}}
\newcommand{\zprobaccelH}{1.0}
\newcommand{\zprimaryjerkH}{5.3\times10^{-2} \pm 1.7\times10^{-1}}
\newcommand{\zsecondaryjerkH}{5.3\times10^{-2} \pm 1.7\times10^{-1}}
\newcommand{\zprobjerkH}{1.0}

\newcommand{\zinitad}{\@m N/A}
\newcommand{\zprimaryaccelad}{1.8\times10^{-1} \pm 2.1\times10^{-1}}
\newcommand{\zsecondaryaccelad}{1.8\times10^{-1} \pm 2.1\times10^{-1}}
\newcommand{\zprobaccelad}{1.0}
\newcommand{\zprimaryjerkad}{1.2\times10^{-1} \pm 2.9\times10^{-1}}
\newcommand{\zsecondaryjerkad}{1.2\times10^{-1} \pm 2.9\times10^{-1}}
\newcommand{\zprobjerkad}{1.0}

\newcommand{\zinitS}{\@m N/A}
\newcommand{\zprimaryaccelS}{1.7\times10^{-2} \pm 1.8\times10^{-2}}
\newcommand{\zsecondaryaccelS}{1.9\times10^{0} \pm 5.5\times10^{-1}}
\newcommand{\zprobaccelS}{120.0}
\newcommand{\zprimaryjerkS}{-2.0\times10^{-2} \pm 2.3\times10^{-1}}
\newcommand{\zsecondaryjerkS}{-2.0\times10^{-2} \pm 2.3\times10^{-1}}
\newcommand{\zprobjerkS}{1.0}

\newcommand{\zinitak}{\@m N/A}
\newcommand{\zprimaryaccelak}{-1.0\times10^{-1} \pm 1.3\times10^{-1}}
\newcommand{\zsecondaryaccelak}{-1.3\times10^{0} \pm 4.4\times10^{-1}}
\newcommand{\zprobaccelak}{7.7}
\newcommand{\zprimaryjerkak}{1.2\times10^{-1} \pm 8.2\times10^{-2}}
\newcommand{\zsecondaryjerkak}{9.4\times10^{-1} \pm 4.3\times10^{-1}}
\newcommand{\zprobjerkak}{3.7}

\newcommand{\zinitac}{\@m N/A}
\newcommand{\zprimaryaccelac}{-2.4\times10^{-1} \pm 6.9\times10^{-2}}
\newcommand{\zsecondaryaccelac}{-1.0\times10^{0} \pm 2.6\times10^{-1}}
\newcommand{\zprobaccelac}{4.3}
\newcommand{\zprimaryjerkac}{-2.0\times10^{-1} \pm 3.1\times10^{-1}}
\newcommand{\zsecondaryjerkac}{-2.0\times10^{-1} \pm 3.1\times10^{-1}}
\newcommand{\zprobjerkac}{1.0}

\newcommand{\zinitQ}{\@m N/A}
\newcommand{\zprimaryaccelQ}{2.0\times10^{-1} \pm 2.7\times10^{-1}}
\newcommand{\zsecondaryaccelQ}{2.0\times10^{-1} \pm 2.7\times10^{-1}}
\newcommand{\zprobaccelQ}{1.0}
\newcommand{\zprimaryjerkQ}{1.1\times10^{-1} \pm 3.8\times10^{-2}}
\newcommand{\zsecondaryjerkQ}{1.7\times10^{0} \pm 7.7\times10^{-1}}
\newcommand{\zprobjerkQ}{2.6}

\newcommand{\zinitE}{\@m N/A}
\newcommand{\zprimaryaccelE}{1.4\times10^{-1} \pm 1.4\times10^{-1}}
\newcommand{\zsecondaryaccelE}{1.4\times10^{-1} \pm 1.4\times10^{-1}}
\newcommand{\zprobaccelE}{1.0}
\newcommand{\zprimaryjerkE}{1.3\times10^{-1} \pm 3.4\times10^{-1}}
\newcommand{\zsecondaryjerkE}{1.3\times10^{-1} \pm 3.4\times10^{-1}}
\newcommand{\zprobjerkE}{1.0}

\newcommand{\zinitT}{\@m N/A}
\newcommand{\zprimaryaccelT}{-6.2\times10^{-1} \pm 2.9\times10^{-1}}
\newcommand{\zsecondaryaccelT}{-6.2\times10^{-1} \pm 2.9\times10^{-1}}
\newcommand{\zprobaccelT}{1.0}
\newcommand{\zprimaryjerkT}{-5.0\times10^{-1} \pm 6.1\times10^{-1}}
\newcommand{\zsecondaryjerkT}{-5.0\times10^{-1} \pm 6.1\times10^{-1}}
\newcommand{\zprobjerkT}{1.0}

\newcommand{\zinitX}{\@m N/A}
\newcommand{\zprimaryaccelX}{-2.7\times10^{-1} \pm 6.3\times10^{-1}}
\newcommand{\zsecondaryaccelX}{-2.7\times10^{-1} \pm 6.3\times10^{-1}}
\newcommand{\zprobaccelX}{1.0}
\newcommand{\zprimaryjerkX}{2.7\times10^{-1} \pm 6.2\times10^{-1}}
\newcommand{\zsecondaryjerkX}{2.7\times10^{-1} \pm 6.2\times10^{-1}}
\newcommand{\zprobjerkX}{1.0}

\newcommand{\zinitA}{\@m N/A}
\newcommand{\zprimaryaccelA}{1.1\times10^{-1} \pm 2.3\times10^{-1}}
\newcommand{\zsecondaryaccelA}{1.1\times10^{-1} \pm 2.3\times10^{-1}}
\newcommand{\zprobaccelA}{1.0}
\newcommand{\zprimaryjerkA}{1.3\times10^{-1} \pm 5.1\times10^{-1}}
\newcommand{\zsecondaryjerkA}{1.3\times10^{-1} \pm 5.1\times10^{-1}}
\newcommand{\zprobjerkA}{1.0}

\newcommand{\zinitD}{\@m N/A}
\newcommand{\zprimaryaccelD}{-9.2\times10^{-1} \pm 3.8\times10^{-1}}
\newcommand{\zsecondaryaccelD}{-9.2\times10^{-1} \pm 3.8\times10^{-1}}
\newcommand{\zprobaccelD}{1.0}
\newcommand{\zprimaryjerkD}{-1.1\times10^{0} \pm 4.3\times10^{-1}}
\newcommand{\zsecondaryjerkD}{-1.1\times10^{0} \pm 4.3\times10^{-1}}
\newcommand{\zprobjerkD}{1.0}

\newcommand{\zinitJ}{\@m N/A}
\newcommand{\zprimaryaccelJ}{1.3\times10^{0} \pm 4.6\times10^{-1}}
\newcommand{\zsecondaryaccelJ}{1.3\times10^{0} \pm 4.6\times10^{-1}}
\newcommand{\zprobaccelJ}{1.0}
\newcommand{\zprimaryjerkJ}{1.2\times10^{0} \pm 4.3\times10^{-1}}
\newcommand{\zsecondaryjerkJ}{1.2\times10^{0} \pm 4.3\times10^{-1}}
\newcommand{\zprobjerkJ}{1.0}

\shorttitle{MCMC Analysis of Terzan 5 Pulsars}
\shortauthors{Prager et al.}

\begin{document}

\title{Using long-term millisecond pulsar timing to obtain physical characteristics of the bulge globular cluster Terzan 5}


\author{Brian~J. Prager\altaffilmark{1},
Scott~M. Ransom\altaffilmark{2},
Paulo~C.C. Freire\altaffilmark{3},
Jason~W.~T. Hessels\altaffilmark{4,5},
Ingrid H. Stairs\altaffilmark{6},
Phil~Arras\altaffilmark{1},
Mario~Cadelano\altaffilmark{7,8}}

\altaffiltext{1}{University of Virginia, Charlottesville, VA 22904, USA}
\altaffiltext{2}{National Radio Astronomy Observatory (NRAO), Charlottesville, VA 22903, USA}
\altaffiltext{3}{Max-Planck-Institut f\"{u}r Radioastronomie Auf dem Hügel 69, D-53121 Bonn, Germany}
\altaffiltext{4}{ASTRON, the Netherlands Institute for Radio Astronomy,Postbus 2, 7990 AA, Dwingeloo, The Netherlands}
\altaffiltext{5}{Anton Pannekoek Institute for Astronomy, University of Amsterdam, Science Park 904, 1098 XH, Amsterdam, The Netherlands}
\altaffiltext{6}{Department of Physics and Astronomy, University of British Columbia,6224 Agricultural Road, Vancouver, BC V6T 1Z1, Canada}
\altaffiltext{7}{Department of Physics and Astronomy, University of Bologna, Viale Berti Pichat 6-2, I-40127, Bologna, Italy}
\altaffiltext{8}{INAF - Bologna Astronomical Observatory, Via Ranzani 1, I-40127 Bologna, Italy}

\begin{abstract}
Over the past decade the discovery of three unique stellar populations and a large number of confirmed pulsars within the globular cluster Terzan 5 has raised questions over its classification. Using the long-term radio pulsar timing of $\Npsr$ millisecond pulsars in the cluster core, we provide new measurements of key physical properties of the system. As Terzan 5 is located within the galactic bulge, stellar crowding and reddening make optical and near infrared observations difficult. Pulsar accelerations, however, allow us to study the intrinsic characteristics of the cluster independent of reddening and stellar crowding and probe the mass density profile without needing to quantify the mass to light ratio. Relating the spin and orbital periods of each pulsar to the acceleration predicted by a King model, we find a core density of $\dencjerk\times$10$^6$ M$_\odot$ pc$^{-3}$, a core radius of $\rcjerk$ pc, a pulsar density profile $n\propto r^{\alphajerk}$, and a total mass of M$_{\rm T}$($R_\perp<$1.0 pc)$\simeq3.0\times$10$^5$ M$_\odot$ assuming a cluster distance of 5.9 kpc. Using this information we argue against Terzan 5 being a disrupted dwarf galaxy and discuss the possibility of Terzan 5 being a fragment of the Milky Way's proto-bulge. We also discuss whether low-mass pulsars were formed via electron capture supernovae or exist in a core full of heavy white dwarfs and hard binaries. Finally we provide an upper limit for the mass of a possible black hole at the core of the cluster of $\mbhupper$.
\end{abstract}

\keywords{globular clusters: individual (Terzan 5), pulsars: general, stars: kinematics and dynamics, black hole physics}

\section{Introduction}

\subsection{Pulsars in Globular Clusters}
Globular clusters (GCs) are extraordinary factories of recycled pulsars; over the last 30 years, 146 pulsars have been discovered in Galactic GCs, with several thousands more still to be discovered \citep{2011MNRAS.418..477B,2015aska.confE..47H}.This pulsar population is completely different from the Galactic population, with a very high fraction of pulsars with rotational periods of a few milliseconds (MSPs). One reason this happens is because GCs are very old stellar systems, so most ``normal'' pulsars in them have long faded into inactive neutron stars (NSs).

Even accounting for the number of inactive NSs, GCs have a very large number of MSPs per unit mass compared to the Galaxy. The reason for this is the high stellar density in the cores of GCs, which permits exchange encounters, and the formation of new binaries consisting of old, inactive NSs with low-mass main sequence stars. The evolution of the latter then causes the system to evolve into a low-mass X-ray binary (LMXB), of which there are many in GCs, and which will likely evolve into an MSP with some sort of companion star, usually a white dwarf (WD) or occassionally a low-mass non-degenerate star. In clusters with very dense cores however, this general evolution could be disturbed by further encounters, generating systems that one would not expect to result from the evolution of LMXBs, such as partially recycled pulsars and secondary exchange encounter products \citep{2014A&A...561A..11V}.

The study of this abundant pulsar population has already resulted in a large number of scientific results, such as NS mass measurements \citep{2003ApJ...584L..13F, 2006A&A...456..295B, 2006ApJ...641L.129C, 2006ApJ...644L.113J, 2012ApJ...745..109L}, tests of general relativity \citep{2006ApJ...644L.113J}, the detection of intracluster gas \citep{2001ApJ...557L.105F}, constraints on the properties of the parent clusters \citep{1993ASPC...50..141P, 1993PhDT.........2A, 2003MNRAS.340.1359F}, and also searches for intermediate mass black holes \citep{2017arXiv170604908F,2017MNRAS.468.2114P,2017Natur.542..203K,2017Natur.542..203K_update}. This last application is especially relevant for this paper.

Among all Galactic GCs, Terzan 5 has the largest number of known pulsars, 36, which represent almost a quarter of the total population of pulsars in GCs\footnote{http://www.naic.edu/~pfreire/GCpsr.html}. The first discovery, J1748$-$2446A, is a very bright pulsar in an eclipsing system \citep{1990Natur.347..650L}, which we would now label as a ``redback''. Ten years later, a second pulsar was announced, J1748$-$2446C \citep{2000MNRAS.316..491L}. In 2004, the commissioning of the S-band receiver at the GBT and the 800-MHz-wide pulsar SPIGOT back-end \citep{2005PASP..117..643K} provided a large increase in sensitivity, which together with better search procedures \citep{2011ascl.soft07017R} resulted in the discovery of 21 new pulsars \citep{2005Sci...307..892R}. Since then another 14 pulsars have been discovered, three of which will soon be reported upon by Cadelano et al. (2017) (J1748$-$2446aj \& J1748$-$2446ak \& J1748-2446al) though only J1748$-$2446aj \& J1748$-$2446ak have timing solutions at this time. This population is quite extraordinary, it includes seven highly eccentric MSP systems, and 5 ``black widow'' and ``redback'' systems, one of which, J1748$-$2446ad, is the fastest-spinning pulsar known \citep{2006Sci...311.1901H}. More details about these ``black widows'' and ``redbacks'' are discussed in Section \ref{subsubsec:bw_rb}.

This uniquely large population of pulsars orbiting in the shared gravitational potential of Terzan 5 can provide a wealth of information about the cluster that is difficult to obtain with optical and infrared observations. Coincidentally, Terzan 5 is one of the most intriguing and least understood GCs in the Galaxy as described in the next subsection.

\subsection{The unusual GC Terzan 5 in the near infrared and X-rays.}

Historically, Terzan 5 has been a very difficult globular cluster to study using ground-based telescopes. Located within the bulge of the Milky Way, it suffers from an average color excess of E(B-V) of 2.38 \citep{1998A&A...333..117B,2007AJ....133.1287V} and strong differential reddening \citep{2012ApJ...755L..32M}. As a result, it has only been through the use of space-based telescopes and adaptive optics in the near infrared that this cluster has become better understood in the past few years. Currently, the most accurate measurement of the cluster distance places Terzan 5 at d=5.9$\pm$0.5 kpc \citep{2007AJ....133.1287V}, the exact value of which can influence the inferred physical characteristics of the system.

\citet{2009Natur.462..483F} uncovered two distinct stellar populations in Terzan 5, one metal rich with [Fe/H]$\simeq$0.3 and a fainter, metal poor population with [Fe/H]$\simeq$-0.2. Additional observations \citep{2013ApJ...779L...5O} of this cluster have revealed a third, even more metal poor component with [Fe/H]$\simeq$-0.79. All of these findings point to the fact that this system has had a unique history and may not be a true globular cluster; instead it may have originally been a dwarf galaxy and undergone tidal stripping due to repeated interactions with the Milky Way's potential \citep{2014MmSAI..85..249M}.

\citet{2009Natur.462..483F} and \citet{2010ApJ...717..653L} give improved measurements of the physical characteristics of the cluster. \citet{2009Natur.462..483F} found the center of gravity (CoG) for Terzan 5 using the absolute positions of stars taken with the European Southern Observatory's Multi-Conjugate Adaptive Optics Demonstrator (ESO-MAD) to be at $\CoGRA$ and $\CoGDec$. \citet{2010ApJ...717..653L} used the observed number density of luminous red giants about the CoG to find a central mass density of (1-4)$\times$10$^6$ M$_\odot$ pc$^{-3}$, a core radius for the red giants of $\lanzonirc$, and a total mass of 2$\times$10$^6$ M$_\odot$, though error estimates are not provided by their work. Using high resolution Hubble Space Telescope data, further studies by \citet{2013ApJ...774..151M} found a core radius of the luminous red giants of $\mioccchirc$.

In terms of MSP formation, Terzan 5 is interesting for a number of reasons. Studies of the radio luminosity function place the number of potentially observable pulsars in the cluster at $\sim$150, which is the largest predicted pulsar population for all known globular clusters \citep{2013MNRAS.431..874C,2013MNRAS.436.3720T}. With 50 X-ray sources identified in the cluster \citep{2006ApJ...651.1098H} and one of the largest stellar collision rates for globular clusters \citep{2003ApJ...591L.131P,2010ApJ...717..653L}, Terzan 5 has a set of physical characteristics that are particularly favorable for the formation of LMXBs and MSPs.

A common concern in previous optical and infrared studies of the structural parameters of Terzan 5 has been finding a way to circumvent the difficulties introduced by the severe reddening and stellar crowding. Pulsar timing allows us to study the mass density of the cluster directly and is, with the exception of needing to determine the distance to the cluster using additional optical data, independent of the reddening. This work accomplishes this by synthesizing and improving upon previously used methods for converting pulsar positions and accelerations into information about the global properties of the system.

\subsection{Constraints on the potential of the cluster from pulsar observations}

As we will see in more detail in Sections \ref{sec:theory} and \ref{sec:psr_to_accel}, the observed variation of the spin and orbital periods ($\dot{P}$ and $\dot{P}_b$) can be used to constrain and measure (respectively) the acceleration produced by the gravitational potential of a spherically symmetric host cluster; the second order spin-period derivative ($\ddot{P}$) can be used to measure the combined ``jerk'' (the time derivative of acceleration) from both the nearby stars and from the overall cluster potential; in this paper we use this newly available information to derive, for the first time, the positions of the pulsars along their lines of sight relative to the center of the cluster. The third order and higher spin-period derivatives are dominated by the presence of nearby stars, however, and do not provide additional constraints on the properties of the cluster as a whole.

These measured accelerations and jerks felt by the pulsars due to the cluster potential can be related to a cluster model through the use of a Markov Chain Monte Carlo (MCMC) sampler. This produces a global measurement of the cluster core density, core radius, spectral index of the pulsar density profile (defined in more detail in Section \ref{sec:initial_fits}), and de-projected positions for each pulsar within the inner few core radii of Terzan 5. Using this method, additional tests can be performed to determine whether a positional offset is needed with respect to the optical CoG of the cluster as well as provide upper limits on the mass of any central black hole in Terzan 5.

As these measurements of the cluster parameters are both reddening independent and calculated using only the gravitating mass of the cluster, we can directly measure the total mass of the cluster at a given radius. This allows us to provide independent measurements of the mass to light ratio that can be used to discuss the most likely formation scenario for the cluster. These measurements also allow us to comment on the dominant mass class within Terzan 5, which can play an important role in understanding the formation mechanism for the large population of MSPs in Terzan 5.

\subsection{Structure of the Paper}

In the remainder of this paper, we begin by discussing the observations and timing campaign of Terzan 5 (Section \ref{sec:data}). We then introduce the models that describe the predicted accelerations and jerks of each pulsar (Section \ref{sec:theory}) and compare these models to globular cluster data produced by the star cluster initializer McLuster \citep{2011MNRAS.417.2300K} (Section \ref{sec:simulations}). Finally, we convert our pulsar timing data into accelerations (Section \ref{sec:psr_to_accel}) and proceed to derive key cluster parameters for Terzan 5 using more traditional timing methods (Section \ref{sec:initial_fits}) as well as with our MCMC models (Sections \ref{sec:MCMC}, \ref{sec:mcmc_47tuc}, and \ref{sec:Results}), including a comparison of each method against previously quoted values in the literature for 47 Tucanae. We finish the paper by discussing our results in the context of what this means for the physical state and origin of Terzan 5 and propose future improvements to our models (Section \ref{sec:discussion}).

\section{Data}
\label{sec:data}

To date there are $\Npsrtot$ known pulsars in Terzan 5, all of which lie within the inner arcminute of the cluster, and the majority within the inner 20 arcseconds. Table \ref{table:Params} gives some of the basic parameters from the timing of these pulsars, including the range of dates used to produce the timing models. Data were taken with the 100-m Robert C. Byrd Green Bank Telescope (GBT) in West Virginia. Observations taken prior to June of 2008 were carried out on the SPIGOT backend \citep{2005PASP..117..643K} and are summarized in the papers by \citet{2005Sci...307..892R} and \citet{2006Sci...311.1901H} along with a discussion of twenty four of the pulsars in our sample. 
Observations taken after this date used the GUPPI (Green Bank Ultimate Pulsar Processing Instrument) \citep{2008SPIE.7019E..1DD} in the coherent search mode. Both observations with SPIGOT and GUPPI were taken with approximately 600 MHz of bandwidth after radio frequency interference (RFI) and band-pass removal and were centered at either 1500 MHz or 2000 MHz.

Search mode observations using coherent dedispersion accumulate full Stokes parameter data in real time before saving the data in a filterbank format. Further details of this process for all $\Npsrtot$ pulsars in Terzan 5 will be presented in an upcoming paper by \citet{Ter5timingprep}. For this work we will only discuss $\Npsr$ of the known pulsars as J1748$-$2446al does not yet have a timing solution.

Of the $\Npsr$ available pulsars for study in the cluster, all have measured first-order spin period derivatives (due, as already mentioned, to the intrinsic spin-down of the pulsars and the acceleration of the cluster potential), 34 have measured second-order spin period derivatives (due in equal parts to nearby stars and the cluster potential), and an additional 13 have a measured third-order spin period derivative or higher.

These higher order spin period derivatives are likely not caused by nearby planets or timing noise. While it is possible that one or two planetary systems are present around our pulsars, the large stellar interaction rate ensures that it is highly unlikely that this is true for the majority of the pulsars. As to the point of the timing noise, MSPs are generally very stable rotators. If we compare our population of MSPs to those in the field of the galaxy, a much smaller fraction have measured second order spin period derivatives and none have a measured third order spin period derivative. Comparing this to the 33 out of 36 timed pulsars in Terzan 5 with a second order spin period derivative (13 of which also have a measured third order spin period derivative) we note that the magnitude of these derivatives can be a few orders of magnitude larger than those seen in the field of the galaxy ($\dot{P}\sim$10$^{-19}$-10$^{-18}$ vs.$\dot{P}\sim$10$^{-21}$-10$^{-20}$)  \citep{2016MNRAS.458.3341D}. We therefore argue that second order and higher spin period derivatives are caused almost exclusively by the presence and movement of nearby stars. Additional arguments for this scenario are given by \citet{1987MNRAS.225P..51B}.

Furthermore, 18 of the $\Npsr$ pulsars are in binaries, 13 of which have measured orbital period derivatives, which in most cases provide direct measurements of the line of sight acceleration of the system, which is dominated by the cluster potential. Combined, we have an unprecedented amount of information from which we can produce a global model for the physical characteristics of the cluster using pulsar timing.

\section{ Theory }
\label{sec:theory}

In this section we derive the accelerations and jerks for a King potential \citep{1962AJ.....67..471K}. For the accelerations we show that the nearest stellar neighbor contribution is negligible, leaving us with a single equation based on the mean-field of the cluster. We discuss the distribution of likelihoods around this mean-field equation in Section \ref{subsubsec:Accel_like}.

\citet{1987MNRAS.225P..51B} showed that the nearest stellar neighbors produce a non-negligible jerk on pulsars in globular clusters. We calculate an analytic solution for the characteristic jerk on a pulsar from all the stars at once to account for this effect. We also calculate the expected distribution of jerks around this characteristic value assuming the inter-stellar distance is much smaller than the size of the cluster.

Finally, we discuss the inclusion of a central black hole in the system and show how this may perturb solutions for the innermost stars. In Section \ref{sec:simulations} we compare our models for the accelerations and jerks to simulated data before applying these methods to Terzan 5.

\subsection{Cluster Geometry}
\label{subsec:geometry}

\begin{figure}[t]
  \centering
  \includegraphics[width=\columnwidth]{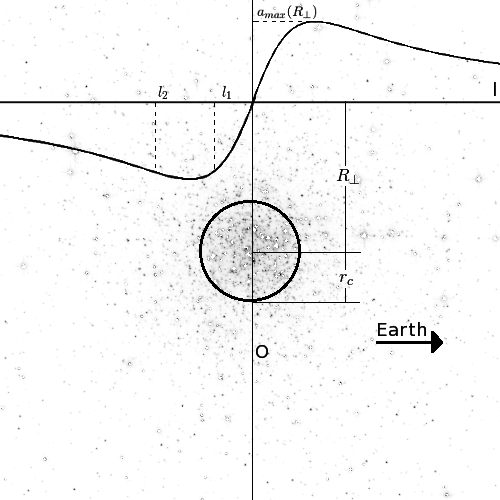}
  \caption
   { Geometry of a pulsar's location with respect to the core of the cluster with an example core radius encircled and the line of sight acceleration over-plotted on top of the optical image of Terzan 5 taken with the Hubble Space Telescope\footnote{https://hla.stsci.edu/}. $O$ is the plane of the sky that runs through the center of gravity for the cluster. $l$ is the line of sight distance out from $O$. $R_\perp$ is the projected separation on the plane of the sky from the center of gravity and $r_c$ is the core radius. For a measured line of sight acceleration below the maximum possible line of sight acceleration $a_{\rm l,max}(R_\perp)$ there are two possible line of sight positions that give the same acceleration, marked by $l_1$ and $l_2$. }
\vspace*{+1mm}
\label{fig:layout}
\end{figure}

We begin by defining a coordinate system for the globular cluster which will be referenced for the rest of this work (Figure \ref{fig:layout}).

The plane running through the CoG and perpendicular to our line of sight (i.e. the plane of the sky) is defined to be $O$. The impact parameter for each pulsar from the CoG is defined to be $R_\perp$. The core radius is defined to be $r_c$, which corresponds to where the mass density drops to a third of its central value. The variables $r_h$ and $r_t$ are the half-mass and the tidal radii of the cluster, respectively.

The line of sight position going perpendicularly through $O$ is defined to be $l$. Positions $l_1$ and $l_2$ represent the two line of sight positions in the cluster where a measured line of sight acceleration $a_l$ may occur. The maximum possible line of sight acceleration at a given value of $R_\perp$ is given by $a_{\rm l,max}(R_\perp)$. The line of sight acceleration rises sharply out to approximately a core radius before declining proportionally to $l/r^3$ out of the plane of the sky.

\subsection{ Acceleration Profile }

In this section we derive the contribution of a pulsar's acceleration that arises due to the globular cluster's mean-field and the nearest neighbors.

\subsubsection{Mean-field Acceleration}
\label{subsubsec:mf_accel}

In order to turn measured accelerations into a probe of the cluster potential we need acceleration as a function of $r_c$, the core density $(\rho_c)$, and the pulsar's spherical radius $(r)$. We begin with a King density profile \citep{1962AJ.....67..471K} for the dominant mass class of the cluster, which most strongly sets the cluster potential:
\begin{equation}
\rho(r)\simeq\rho_c\left[1+\left(\frac{r}{r_c}\right)^2\right]^{^-\frac{3}{2}}.
\label{eq:king_cum_rho}
\end{equation}

Integrating this equation radially yields the interior mass at any given radius $r$. We multiplied this by $-G/r^2$ to obtain a general form of the cluster acceleration felt at any given radius out from the core:
\begin{equation}
a_{\rm r}(r) = -4\pi G\rho_cr_c^3r^{-2}\left[\sinh^{-1}\left(\frac{r}{r_c}\right)-\frac{r}{r_c\sqrt{1+\left(r/r_c\right)^2}}\right]
\label{eq:king_amax_radial_rho}
\end{equation}
where $a_{\rm r}(r)$ is the predicted radial mean-field acceleration and was derived for the first time explicitely by \citet{2005ApJ...621..959F}.

Projecting the acceleration along our line of sight by a factor of $l/r$ and substituting typical values for globular clusters like Terzan 5, Equation \ref{eq:king_amax_radial_rho} reduces to:
\begin{equation}
\begin{split}
a&_{\rm l}(l,r) = -3.5\times10^{-7}\left(\frac{\rho_c}{10^6\,{\rm M}_\odot\,{\rm pc}^{-3}}\right)\left(\frac{l}{0.2\,{\rm pc}}\right) \\
&\times \left(\frac{r}{r_c}\right)^{-3}\left[\sinh^{-1}\left(\frac{r}{r_c}\right)-\frac{r}{r_c\sqrt{1+\left(r/r_c\right)^2}}\right] {\rm m\,s^{-2}}\,,
\end{split}
\label{eq:king_amax}
\end{equation}
where we define $a_l=a_{\rm l}(l,r)$ to be the predicted mean-field line of sight acceleration for the rest of this paper. The shape of this distribution is shown in Figure \ref{fig:layout}.

\subsubsection{Nearest Neighbor Accelerations}

\citet{1943RvMP...15....1C} derives the {\it Holtzmark} probability distribution for acceleration due to an infinite distribution of stars with a given mean density. In Appendix \ref{appendix:nn_accel}, we use this method to determine the probability that a nearest-neigbor can produce an acceleration equivalent to, or greather than, the mean field.

The cumulative probability ($P_{\rm c}$) that the nearest neighbor acceleration $a_{\rm NN}$ is larger than a value $a_l$ is given by Equation \ref{eqn:appendix_cum_prob}:
\begin{equation}
P_{\rm c}(a_l) = \int_{a_l}^\infty da_l' P(a_l') = \frac{1}{\sqrt{2\pi} }\left( \frac{a_{\rm NN}}{ \left| a_l \right|} \right)^{3/2}\,.
\label{eqn:p_l_greater}
\end{equation}

In Section \ref{subsec:sim_accel} we will compare our predicted probability to the mean-field acceleration using simulated cluster data in order to show that the acceleration due to the nearest stellar neighbor a is a negligibly small effect in our models.

\subsection{ Jerk Profile }

\subsubsection{Converting $\ddot{P}$ to Jerks}

Due to the motion of a pulsar through its potential relative to the observer, a pulsar with a rest frame period of $P_0$ is observed to have a period P of

\begin{equation}
P = \left[1+(\bvec{V}_p-\bvec{V}_{\rm bary})\cdot\bvec{n}/c\right]P_0\;,
\label{eq:observed_P}
\end{equation}
where $\bvec{V}_p$ is the pulsar velocity, $\bvec{V}_{\rm bary}$ is the velocity of the solar system barycenter, and $\bvec{n}$ is the unit vector along our line of sight.

Following the prescription of \citet{1993ASPC...50..141P} the derivative of Equation \ref{eq:observed_P} yields expansions of the spin-period derivatives that can be expressed as an acceleration or its time derivatives. Limiting ourselves to a pulsars jerk, this becomes
\begin{equation}
\frac{\ddot{P}}{P}=\frac{1}{c}\dot{a}\cdot\bvec{n}\,,
\label{eq:pdot_eqn}
\end{equation}
which allows us to derive the pulsar's jerk using the second order derivative of the pulsar acceleration, $\ddot{P}$.

\subsubsection{Mean-field Jerks}

From \citet{1993ASPC...50..141P} we use Equation 4.3 for defining the mean-field jerk felt by a pulsar within the inner few core radii of the cluster:
\begin{equation}
    \left(\frac{\ddot{P}}{P}\right)_{\rm mf} = \dot{\bvec{a}}_{\rm mf}\cdot\frac{\bvec{n}}{c}=\frac{4}{3}\pi G\rho(r)\frac{v_l(r)}{c}\,,
\label{eqn:jerk_mf}
\end{equation}
where $v_l(r)$ is the line-of-sight velocity of the pulsar.

In Section \ref{subsec:sim_jerk} we will compare the distribution of measured mean-field jerks from simulations to Equation \ref{eqn:jerk_mf}.

\subsection{Nearest Neighbor Jerks}

The characteristic jerk experienced at the location of the pulsar due to the surrounding stars is derived in Appendix \ref{appendix:jerks} and is given by the following formula:
\begin{equation}
\dot{a}_0 = \frac{2\pi \xi}{3} G \langle m \rangle \sigma n\,,
\label{eqn:char_jerk}
\end{equation}
where $\dot{a}_0$ is the characteristic jerk and $\xi$ is a numerical factor that arises from integrating over the Maxwell-Boltzmann distribution in the vicinity of the pulsars and can be approximated by $\xi\simeq$3.04. The variables $m$, $\sigma$, and $n$ are the mass of the nearby stars, the velocity dispersion of the nearby stars, and the number density of stars near the pulsar, respectively.

\subsection{Central Black Hole}
\label{sec:theory_bh}

We tested for the presence of an intermediate mass black hole (IMBH) in Terzan 5 by perturbing the standard King density profile assuming a black hole is fixed to the center of the cluster. In general, this assumption is most valid for larger black hole masses, since variations in the central cluster potential will perturb the black hole away from the center for smaller black hole masses.  Our assumption should be valid, though, when determining an upper mass limit for a central black hole.

We begin by noting that for a given mass, the black hole will have a radius of influence ($r_i$) given by Equation 3 from \citet{2004ApJ...613.1133B}
\begin{equation}
r_i=\frac{3M_{\rm BH}}{8\pi\rho_cr_c^2}\,,
\end{equation}
where $M_{\rm BH}$ is the mass of the central black hole.

At $r_i$ and beyond, the density profile follows the standard King model for globular clusters \citep{2004ApJ...613.1143B}. Within this radius the density profile obeys the following formula:
\begin{equation}
\rho_{\rm BH}\propto r^{-1.55}\,,
\end{equation}
where our selection of a density profile following a $-1.55$ power law is taken from \citet{2004ApJ...613.1143B}, where they found through N-body simulations that a system with multiple component masses in the core follows this density profile. This is close to the results of \citet{1976ApJ...209..214B}, which find that for a more top-heavy initial mass function (IMF) around a black hole, the density profile must scale as $r^{-1.5}$.

Using this modified density profile within the radius of influence, we derive a perturbed model equation:
\begin{equation}
a_l(l,r)=\frac{4\pi G}{r^2}\frac{l}{r}\left[\int_0^{r_i} r^2\rho_{\rm BH}dr+\int_{r_i}^r r^2\rho_{\rm King}\right]\,,
\label{eqn:model_accel_bh}
\end{equation}
where $\rho_{\rm King}$ is the King model for density in Equation \ref{eq:king_cum_rho}.

Allowing the black hole mass to vary in our simulation allows the pulsars to potentially enter its the radius of influence, drastically altering their modeled acceleration. 

The expected posterior distribution should therefore show that for small black hole masses the black hole is not massive enough to produce a measurable acceleration at the position of a pulsar. As the black hole mass grows, so does the chances of strongly perturbing a pulsar, thereby giving better sensitivity to the likelihood of having a black hole in the system.

For the purposes of this work, we do not consider the jerk produced by the black hole, though we do consider the jerk felt by each pulsar due to the modified density profile in the core.

\section{ Cluster Simulations }
\label{sec:simulations}

\begin{figure}[t]
  \centering
  \includegraphics[width=\columnwidth]{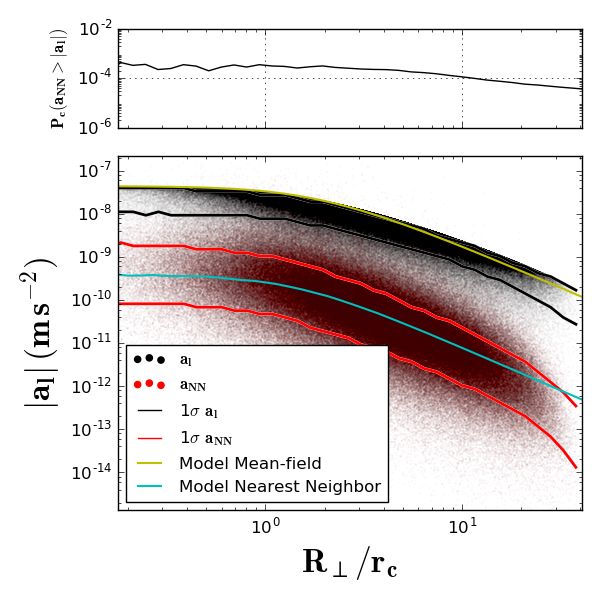}
  \caption
   { Measured line of sight accelerations within cylindrical shells for a simulated Terzan 5 cluster with $M_t$=2$\times$10$^5$ M$_\odot$ and a $r_c$=0.2 pc showing that nearest neighbor accelerations matching or exceeding the mean-field value are rare. The black points show the measured mean-field acceleration and the red points show the measured nearest neighbor values. The black and red lines show the 1-$\sigma$ confidence intervals for the mean-field and the nearest neighbors respectively. Overplotted are the model accelerations given by Equations \ref{eq:king_amax} and \ref{eq:aNN}. The upper plot shows the probability of the measured nearest neighbor acceleration exceeding the mean-field according to Equation \ref{eqn:p_l_greater}. }
\vspace*{+1mm}
\label{fig:median_accel}
\end{figure}

Using simulations, we verified the results of Section \ref{sec:theory} as well as those originally presented by \citet{1993ASPC...50..141P} and \citet{1993PhDT.........2A}. We accomplished this using the C-based artificial star cluster initializer, McLuster \citep{2011MNRAS.417.2300K}, which allows us to create star clusters with a variety of different cluster potentials, masses, ages, and various other structural parameters. In order to simulate Terzan 5, we used the results of \citet{2013ApJ...774..151M} to set the cluster potential to a King model with $r_{\rm h}$=0.98 pc and a concentration parameter of $W_0$=7.2.

We used the Kroupa initial mass function (IMF) \citep{2001MNRAS.322..231K} for our simulations, which has a slope of $\alpha=-1.3$ over the mass ranges $0.08\,{\rm M}_\odot\leq m\leq0.5\,{\rm M}_\odot$ and a slope of $\alpha=-2.3$ for masses $0.5\,{\rm M}_\odot\leq m\leq100\,{\rm M}_\odot$. We then evolved all the stars in the cluster in place for 12 Gyr to create data consistent with the oldest population in Terzan 5 \citep{2013ApJ...779L...5O}.

For our models we varied the total mass of the cluster $M_t$, $r_t$, and $r_c$ while holding all other cluster parameters constant. To ensure that models of the same mass but different limiting radii were consistent, we used a single seed to hold constant the stellar masses of individuals stars drawn from the IMF as we varied the cluster parameters.

\begin{figure}[t]
  \centering
  \includegraphics[width=\columnwidth]{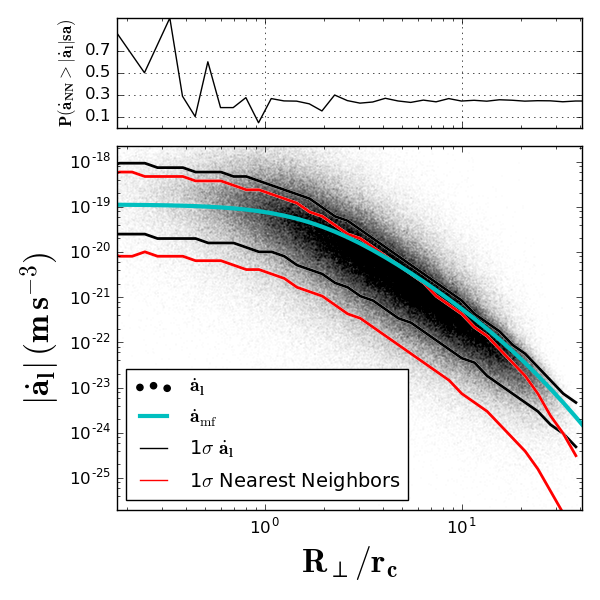}
  \caption
   { Measured line of sight jerks within cylindrical shells for a simulated Terzan 5 with $M_t$=2$\times$10$^5$ M$_\odot$ and $r_c$=0.2 pc. The black points show the combined pair-wise jerk from all the stars in the simulation for any given particle. The black and red lines show the 1-$\sigma$ confidence intervals for the combined and nearest neighbor values respectively. Overplotted is the mean-field jerk given by Equation \ref{eqn:jerk_mf}. The upper plot shows the probability of the measured nearest neighbor jerk exceeding the mean-field jerk within each cylindrical bin. }
\vspace*{+1mm}
\label{fig:median_jerk}
\end{figure}

We then output a list of positions and velocities for each star from the simulation and calculated the acceleration and jerk due to the collection of all the particles in the simulation in order to test our model equations. The details of these calculations are discussed in Sections \ref{subsec:sim_accel} and \ref{subsec:sim_jerk}.

\subsection{ Pulsar Accelerations}
\label{subsec:sim_accel}

Using McLuster, we measured the mean-field acceleration on each particle in the simulation, as well as contributions from the nearest neighbors. For the mean-field we calculated the mass interior to the radial position of each particle to compare to our model acceleration given by Equation \ref{eq:king_amax}. The nearest neighbor accelerations were calculated for each particle by finding its nearest companion and calculating its acceleration directly.

We find that for all values of $M_t$, $r_c$, and $r_t$ that would be applicable to a Terzan 5 like system, the accelerations are dominated by the mean-field. We present our results for one such cluster simulation with $M_t$=2$\times$10$^5$ M$_\odot$ and $r_c$=0.2 pc as these simulation parameters are close to those derived in Section \ref{sec:Results} while still having few enough particles to model efficiently.

Figure \ref{fig:median_accel} shows the measured values of $a_l$ and $a_{NN}$ for one of the Terzan 5 simulations in equally log-spaced cylindrical shells centered around the CoG. From this figure we estimate the probability of a nearest neighbor being the dominant source of acceleration for a pulsar according to Equation \ref{eqn:p_l_greater}. 

For an acceleration $a_{\rm l}$=10$^{-7}$ m s$^{-2}$ and a nearest neighbor acceleration of $a_{\rm NN}\sim10^{-9}$ m s$^{-2}$, we find $P_c(a_{\rm l})\sim10^{-4}$. This agrees with the simulated data as shown in the top panel of Figure \ref{fig:median_accel}, which shows that the nearest neighbors represent a negligible contribution to the acceleration felt by any single star. We therefore consider it unlikely that nearest neighbors can account for the accelerations felt by each pulsar and do not consider their effect in our calculations.

\subsection{ Pulsar Jerks }
\label{subsec:sim_jerk}

Figure \ref{fig:median_jerk} shows the measured jerk due to the pair-wise interaction of all particles in the simulation ($\dot{a}_l$) as well as the 1-$\sigma$ region containing the measured nearest neighbor induced jerks within log-spaced cylindrical shells for a simulated Terzan 5 cluster. The top panel shows the ratio of measured nearest neighbor jerks greater than, or equal to, the mean-field jerk.

\begin{figure}[t]
  \centering
  \includegraphics[width=\columnwidth]{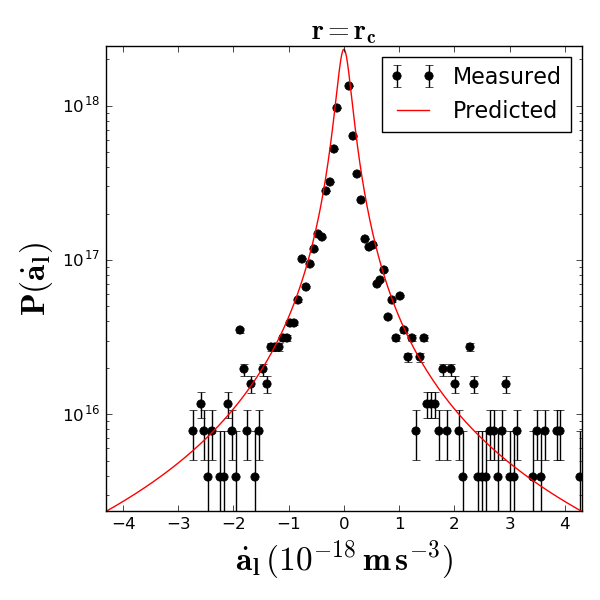}
  \caption
   { Measured jerks within a spherical shell centered about the core of a simulated Terzan 5 cluster with $M_t$=2$\times$10$^5$ M$_\odot$ and $r_c$=0.2 pc. The data points show the measured probability of finding a given value of jerk along with its error for bins with greater than 10 counts. The red line shows the calculated nearest neighbor PDF from Equation \ref{eqn:jerk_pdf}. }
\label{fig:measured_theory_jerk_pdf}
\end{figure}

We find that the nearest neighbor jerks are always of a similar magnitude as the mean-field measurement, regardless of the cluster's $M_t$, $r_c$, and $r_t$. We have plotted the mean-field jerk defined by Equation \ref{eqn:jerk_mf} in cyan and find good agreement with the data. We present our results for one such cluster simulation with $M_t$=2$\times$10$^5$ M$_\odot$ and $r_c$=0.2 pc as these simulation parameters are close to those derived in Section \ref{sec:Results} while still having few enough particles to model efficiently.

From Appendix \ref{appendix:jerks} the distribution of possible line of sight jerks ($\dot{a}_l$) for the pulsar due to its nearest neighbor is given by the following Lorentzian formula:
\begin{equation}
P(\dot{a}_l) = \frac{\dot{a}_0/\pi}{\dot{a}_l^2 + \dot{a}_0^2}\,.
\label{eqn:jerk_pdf}
\end{equation}

Figure \ref{fig:measured_theory_jerk_pdf} shows the measured probability density function (PDF) for the distribution of nearest neighbors in a single spherical shell for a simulated Terzan 5 cluster. Over-plotted is the predicted PDF for pulsar jerks given by Equation \ref{eqn:jerk_pdf}. This derived PDF has good agreement across all cluster masses and radii for spherical shells out to intermediate radii ($r\lesssim3.0r_c$). Beyond these radii we begin to see deviations in the tail of the distribution, which is likely due to the stars no longer following a lowered Maxwellian distribution.

When performing our final analysis we do not consider the jerks of those pulsars with $r>3.0r_c$, though we still use the acceleration information for these pulsars to inform us about the outer parts of the cluster.

\section{Converting Pulsar Timing to Accelerations}
\label{sec:psr_to_accel}

\subsection{Isolating Cluster Acceleration}
\label{subsec:isolating_cluster_spindown}

Before studying the cluster potential further, we define our method for converting pulsar timing data into an acceleration. To do so, we follow the prescription of \citet{1993ASPC...50..141P} and \citet{1993PhDT.........2A}, which relates changes in the pulsar's spin and orbital periods to an acceleration.

For a pulsar sitting in a smooth, spherically symmetric potential the observed change in a pulsar's spin period is the result of the intrinsic spin down of the pulsar and any additional acceleration along our line of sight. The measured spin period changes can therefore be decomposed into the following form:
\begin{equation}
\frac{a_{\rm meas}}{c} \equiv \left(\frac{\dot{P}}{P}\right)_{\rm meas} = \left(\frac{\dot{P}}{P}\right)_{\rm int} + \frac{a_c}{c} + \frac{a_g}{c} + \frac{a_s}{c} + \frac{a_{\rm DM}}{c}\,,
\vspace*{+2mm}
\label{eq:pdot_meas_terms}
\end{equation}
where {\scriptsize $\left(\dot{P}/P\right)_{\rm int}$} is the spin period change associated with the intrinsic pulsar spin-down, $a_c$ is the line of sight acceleration due to the globular cluster potential on a pulsar, $a_g$ is the acceleration due to the Galactic potential, $a_s$ is the apparent acceleration from the Shklovskii effect \citep{1970SvA....13..562S}, and $a_{\rm DM}$ is the apparent acceleration due to errors in the changing dispersion measure (DM) towards the pulsar.

\begin{figure}[t]
  \centering
  \includegraphics[width=\columnwidth]{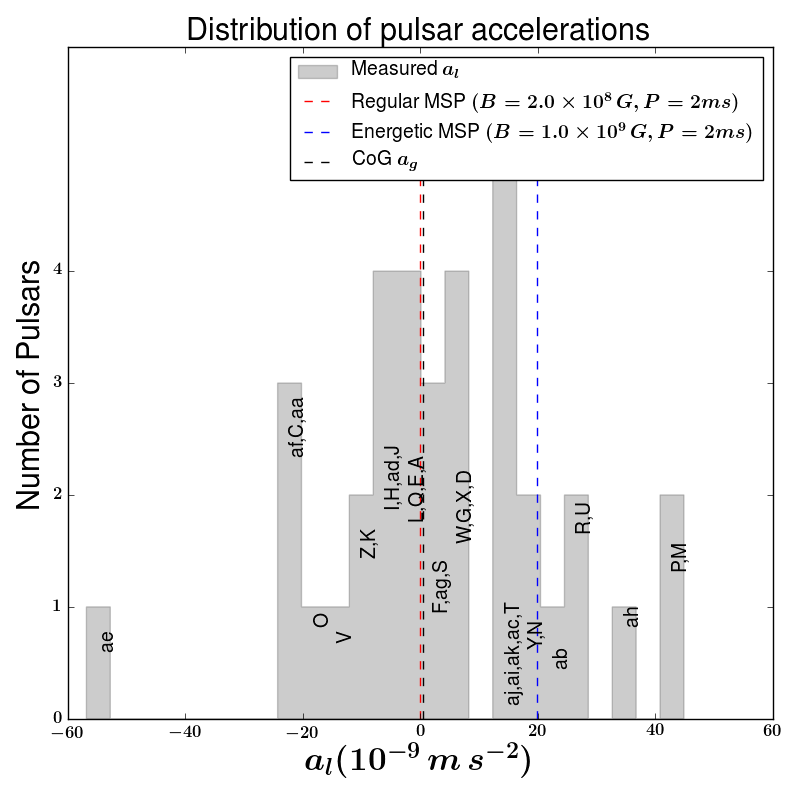}
  \caption
   { Histogram of the measured accelerations $a_{\rm l}$ for Terzan 5 pulsars. Over-plotted in red is a typical MSP spin-down shown as an apparent acceleration and in blue is an example of an energetic MSP's spin-down shown as an apparent acceleration. Lying nearly coincident with the acceleration for the spin-down of a typical MSP is the acceleration due to the Galactic potential on the CoG of the cluster, which is shown in black and is largely overlapped by the red line. For all but the most energetic pulsars the cluster potential is the dominant source of acceleration. }
\label{fig:measured_psr_accels}
\end{figure}

Figure 5 shows the measured acceleration for each pulsar in Terzan 5 along with the predicted accelerations one would measure for different intrinsic spin-down rates of a pulsar; additionally, the expected acceleration due to the Galactic potential at the core of the cluster is plotted as a dashed black line.

\subsubsection{Accelerations from intrinsic spin-down}
\label{subsubsec:intrinsic_spindown}

We calculated the intrinsic spin-down by relating the magnetic field strength of the pulsars to their spin period derivative for isolated pulsars and those binary systems without a measured orbital period derivative. In Section \ref{subsec:ac_binaries} we discuss how we can circumvent the need to measure the intrinsic spin-down of some of our pulsar binaries. 

For a simple model of a pulsar with magnetic dipole emission and a braking index of n=3, the $\dot{P}$ for a typical MSP is given by:
\begin{equation}
\resizebox{.9\columnwidth}{!}{$c\left(\frac{\dot{P}}{P}\right)_{\rm int} = 7.96\times10^{-10}\left(\frac{B}{2\times 10^8 \,{\rm G}}\right)^2\left(\frac{2 \,{\rm ms}}{P}\right)^2 \,{\rm m}\,{\rm s}^{-2}\,,$}
\label{eq:spindown}
\end{equation}
where $B$ is the surface magnetic field strength of the pulsar and the numerical factor has been scaled to appropriate values for an MSP.

Magnetic field strengths for pulsars are typically measured using the pulsar's $P$ and $\dot{P}$ values. Since pulsars in globular clusters have measured values of $\dot{P}$ that are dominated by the cluster potential however, measurements of magnetic field strengths for these systems have not been extensively studied to date.

\begin{figure}[t]
  \centering
  \includegraphics[width=\columnwidth]{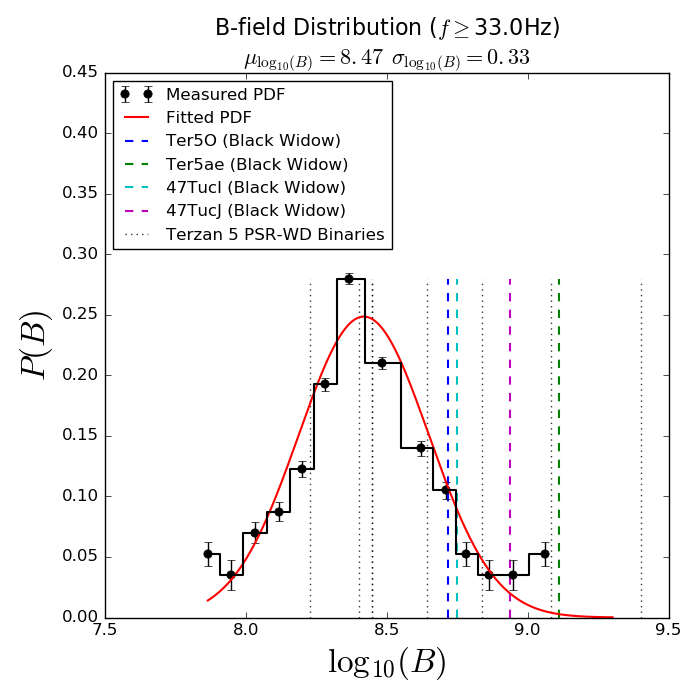}
  \caption 
   { Measured distribution of magnetic field strengths for millisecond pulsars with a spin frequency greater than 33 Hz and magnetic field strengths less than 1.5$\times$10$^9$ G in the field of the Galaxy. Over-plotted is a log-normal fit to the data used to draw a log-likelihood of finding a pulsar with a given magnetic field strength. Vertical lines show the measured magnetic field strength of globular cluster black widows (dashed lines) and pulsar-white dwarf binaries in Terzan 5 (dotted lines) after using the measured $\dot{P}_b$ to remove the cluster induced spin-down as shown in Equation \ref{eq:meas_pbdot_pb}. }
\vspace*{+1mm}
\label{fig:atnf_bfield_dist}
\end{figure}

To get around this problem we used the known magnetic field strengths for similar Galactic MSPs taken from the ATNF catalog\footnote{http://www.atnf.csiro.au/people/pulsar/psrcat/} \citep{2005yCat.7245....0M}. These values of magnetic field strength were then fit to a log-normal PDF. We have plotted the results of these fits for pulsars with a spin frequency greater than 33 Hz and magnetic field strengths less than 1.5$\times$10$^9$ G as a function of $\log_{10}(B)$ in Figure \ref{fig:atnf_bfield_dist}. We find that the median and standard deviation of the resulting normal distribution are given by $\mu_{\log_{10}(B)}=8.47$ and $\sigma_{\log_{10}(B)}=0.33$.

In Section \ref{sec:MCMC} we will discuss how Equation \ref{eq:spindown} and the log-normal fit shown in Figure \ref{fig:atnf_bfield_dist} are used to fit for cluster parameters in our MCMC sampler.

\subsubsection{Accelerations from the Galactic potential}

Next we calculate the acceleration due to the Galactic potential for each pulsar. To do this, we used a distance for the Sun from the Galactic center of $R_0=8.34\pm0.16$ kpc and a rotational speed of the Galaxy at the Sun of $\Theta_0=240\pm8$ km/s, as measured by modeling of the Radial Velocity Experiment and the Geneva-Copenhagen Survey  \citep{2014ApJ...793...51S}. For a flat rotation curve and a distance to Terzan 5 of $d=5.9\pm0.5$ kpc, Equation 5 from \citet{1995ApJ...441..429N} gives the acceleration due to the differential Galactic rotation at the location of each pulsar as observed from the Earth:
\begin{equation}
a_g\cdot \bvec{n} = -\cos(b)\left(\frac{\Theta_0^2}{R_0}\right)\left(\cos(l)+\frac{\beta}{\sin^2(l)+\beta^2}\right)\,{\rm m}\,{\rm s^{-2}}\,,
\end{equation}
where $\beta=(d/R_0)\cos(b)-\cos(l)$. 

At the CoG for Terzan 5 ($\CoGl$, $\CoGb$) we find an acceleration of $\ag$. This is shown in Figure \ref{fig:measured_psr_accels} as the dashed black line, which is much smaller than the observed apparent accelerations and is nearly coincident with the plotted apparent acceleration of a typical MSP spin-down rate.

\subsubsection{Accelerations from Shklovskii effect}
Next we estimated the Shklovskii effect at the distance of Terzan 5, which is an apparent acceleration due to the proper motion of the pulsar. From \citet{1970SvA....13..562S} we have:
\begin{equation}
a_{s} = 4.29 \times 10^{-12}\left(\frac{d}{5.9\,{\rm kpc}}\right)\left(\frac{\mu_T}{{\rm mas}\,{\rm yr}^{-1}}\right)^2\,{\rm m}\,{\rm s}^{-2}\,,
\label{eq:shk}
\end{equation}
where $d$ is the cluster distance and $\mu_T$ is the proper motion.

As Terzan 5 is near the center of the Galaxy, we used the tangential velocity of the Sun with respect to the center of the Galaxy, $\Theta_0$, as a representative value for the system. The proper motion is therefore given by:
\begin{equation}
\mu_T = 8.5 \left(\frac{\Theta_0}{240 {\rm km s}}\right)\left(\frac{5.9 \:{\rm kpc}}{d}\right) \:{\rm mas}\,{\rm yr}^{-1}\,.
\end{equation}

Using Equation \ref{eq:shk} we find $a_{s}\sim 4.2 \times 10^{-12}$ m s$^{-2}$, which is much smaller than the other sources of acceleration. Given the relative scale of this term, the Shklovskii effect is a very tiny contribution to the measured acceleration for any reasonable value of transverse velocity.

\subsubsection{Accelerations from DM errors}

Propagating radio waves in the ionized interstellar medium undergo delays in their arrival time at Earth due to dispersion. For a pulsar this means that the arrival time of the pulse is delayed as a function of the observing frequency:
\begin{equation}
\Delta t_{\rm DM}=1.85\left(\frac{DM}{\text{pc}\,\text{cm}^{-3}}\right)\left(\frac{\nu}{1500\,\text{MHz}}\right)^{-2}\:\text{ms}
\label{eq:dm_delay}
\end{equation}
where $\Delta t_{\rm DM}$ is the dispersive delay time, the DM is the dispersion measure of the pulsar and is related to the number of free electrons between us and the pulsar, and $\nu$ is the observing frequency.

If there are unaccounted for errors in the DM as a function of time ($\Delta{\rm DM}(t)$), it is possible to produce apparent accelerations in the pulsar's timing. In the upcoming \citet{Ter5timingprep} paper, the observed DM errors were measured to be $\Delta{\rm DM}$ $\lesssim$ 10$^{-2}$ pc cm$^{-3}$ for all pulsars as observed over a 10 year time period. Using this as an upper limit on the possible values of $\Delta{\rm DM}(t)$ at an observing frequency of 1500 MHz, Equation \ref{eq:dm_delay} gives an excess delay of $\Delta t_{\rm DM}\lesssim$18 $\mu$s for our uncertainty in DM over the past ten years of timing.

We convert the observed change in arrival time of a given pulse into a frequency derivative by looking at the change in the phase of the pulse over a given timescale:
\begin{equation}
\dot{f} = 2\left(\frac{\Delta t_{\rm DM}}{PT^2}\right)\,,
\label{eq:Pdot_dm}
\end{equation}
where T is the timescale over which we measure the error in DM.

Converting this into a period derivative, the acceleration from the stochastic DM error is given by:
\begin{equation}
a_{\rm DM} = -2c\left(\frac{\Delta t_{\rm DM}}{T^2}\right)\,,
\label{eq:accel_dm_no_norm}
\end{equation}
which can be expressed in more convenient units of:
\begin{equation}
a_{\rm DM} = -6.1\times10^{-14}\,\left(\frac{\Delta t_{\rm DM}}{1\,\mu\text{s}}\right)\left(\frac{10\,\text{yrs}}{T}\right)^2\,.
\label{eq:accel_dm}
\end{equation}

For a delay of $\Delta t_{\rm DM}$=18 $\mu$s due to unexplained DM variations over a baseline of T=10 years, the apparent acceleration is $a_{\rm DM}\sim10^{-13}$ m s$^{-2}$. This is a very minor contribution to our measured acceleration and as such is not corrected for in our data. Also, as it is very unlikely for the DM variations to be monotonic over this span of time, the actual contribution to the measured acceleration of each pulsar from the DM is likely much smaller than this value \citep{2013MNRAS.429.2161K}.

These measurements of acceleration from intrinsic spin-down, the Galactic potential, the Shklovskii effect, and the error in DM place strong constraints on the allowed range of accelerations any cluster model may produce. In the next few sections we will discuss how we used the timing of pulsar binaries to provide even better constraints on our models by avoiding any assumptions of intrinsic pulsar spin-down rates.

\subsection{Measuring $a_c$ using pulsar binaries}
\label{subsec:ac_binaries}

In addition to the spin periods of each pulsar, we also used the orbital period ($P_b$) and orbital period derivatives ($\dot{P}_b$) of pulsar binaries to probe the cluster potential. Each of the thirteen systems with a measured value of $\dot{P}_b$ appear to be cluster dominated, and do not experience any measurable orbital decay due to general relativistic effects \citep{1985AIHS...43..107D}. Without an intrinsic effect within the binary system causing the orbital period derivative, this means changes in the orbital period are almost entirely due to the cluster potential.

We have already shown that the Galactic rotation and observed Shklovskii effect are small and can be well accounted for in our timing. We therefore may express the orbital period derivative in a similar form as Equation \ref{eq:pdot_meas_terms}
\begin{equation}
\left(\frac{\dot{P}_b}{P_b}\right)_{meas} = \frac{a_c}{c} + \frac{a_g}{c} + \frac{a_s}{c}\,.
\label{eq:meas_pbdot_pb}
\end{equation}

In the next section, we discuss briefly a subclass of millisecond pulsars with measurable orbital period derivatives that must be carefully examined before applying Equation \ref{eq:meas_pbdot_pb} to their timing results.

\subsubsection{Black Widow \& Redback Pulsars}
\label{subsubsec:bw_rb}

Of the thirteen pulsar binaries with a measured $\dot{P}_b$, eight are in binaries with a white dwarf companion and are well characterized by Equation \ref{eq:meas_pbdot_pb}. The remaining five pulsars are in a binary system with an companion star that is not fully degenerate. These systems are referred to as black widows if they have a companion star less massive than $\sim$0.08 M$_\odot$ or a redback if it is more massive than $\sim$0.08 M$_\odot$. An important characteristic of these systems is that some experience stochastic changes in their orbital parameters which could manifest as an apparent acceleration in the timing data.

While some black widow systems do not show orbital phase variations, all redbacks systems discovered to date show changes in their orbital properties that are on average an order of magnitude larger than any seen in black widows. Even over a few year time scale the timing of a redback becomes nearly intractable as the stochastic wander of orbital properties evolve.

Within Terzan 5 there are three redbacks (J1748$-$2446A, J1748$-$2446P, and J1748$-$2446ad) and two black widows (J1748$-$2446O and J1748$-$2446ae). Due to the severity of the orbital phase changes seen in redbacks, we do not use the measured orbital period derivatives for these three systems. As for the two black widows (J1748$-$2446ae and J1748$-$2446O), with measured orbital period derivatives, we motivate our inclusion of these two pulsars with two different arguments.

The first argument is that observed variations in the orbital parameters of black widows and redbacks are believed to be stochastic or quasi-periodic in nature \citep{1994ApJ...436..312A}. It is therefore unlikely that the changes in the orbital period would be sufficiently monotonic to produce a large apparent acceleration for these systems. The second argument is that if either of these systems has an intrinsic spin-down rate after subtraction of the apparent cluster acceleration that is not typical of MSPs, then that may be indicative of a timing model influenced by changing orbital properties.

Table \ref{table:bws} shows the apparent line of sight accelerations for eight black widows using the measured values of $\dot{P}_b$ and Equation \ref{eq:meas_pbdot_pb}. Of these eight systems, four have been confirmed to have orbital phase wander. These systems show line of sight accelerations in excess of $10^{-7}$ m s$^{-2}$, whereas the remaining black widows all have accelerations ranging between 5$\times$10$^{-9}$ and 5$\times$10$^{-8}$ m s$^{-2}$. This may imply that only a handful of black widows are in a state where the mechanism that drives orbital wander is enhanced.

We also confirmed that the measured values of $\dot{P}_b$ are cluster-induced by looking at the energetics of the system. Taking the difference between the measured acceleration given by Equation \ref{eq:pdot_meas_terms} and the binary acceleration given by Equation \ref{eq:meas_pbdot_pb} we are left with the intrinsic spin-down component of the pulsar in the binary system:
\begin{equation}
\left(\frac{\dot{P}}{P}\right)_{\rm int} = \left(\frac{\dot{P}}{P}\right)_{\rm meas} - \left(\frac{\dot{P}_b}{P_b}\right).
\label{eq:ae_o_pdot}
\end{equation}

Table \ref{table:bws} shows the magnetic field strength for the inferred intrinsic spin down rate of each black widow for which a $\dot{P}_b$ value has been measured in 47 Tucanae and Terzan 5. Figure \ref{fig:atnf_bfield_dist} shows the inferred magnetic field strength of each of these black widows as compared to the observed magnetic field distribution in the field of the Galaxy as recorded by the ATNF catalog \citep{2005yCat.7245....0M}. Also plotted are the inferred magnetic field strengths of regular pulsar-white dwarf binaries. The The black widows appear to have typical magnetic field strengths and therefore have cluster dominated orbital period derivatives. 47Tuc0 is the only exception to this, which has been confirmed to have orbital phase wander and as such produces an unphysical magnetic field strength, implying the measured value of $\dot{P}_b$ for this system is dominated by the physics of its orbital phase wander. These results show that the energetics of the two black widows J1748$-$2446O and J1748$-$2446ae appear to be cluster dominated and do not have an excess in their measured values of $\dot{P}_b$ from phase wander. 

For the rest of this paper, we will assume that the orbital period derivatives of both J1748$-$2446O and J1748$-$2446ae are dominated by the cluster potential, and will use the measured values of $\dot{P}_b$ for each to help constrain the cluster parameters.

\section{Traditional Parameter Fits Using Pulsar Timing}
\label{sec:initial_fits}

In this section we used the methods of \citet{1993ASPC...50..141P}, \citet{1993PhDT.........2A}, \citet{2001ApJ...557L.105F}, and \citet{2003MNRAS.340.1359F} to relate the on-sky position of a pulsar and the predicted cluster parameters of the host system. We will use this method to find an initial estimate of the Terzan 5 cluster characteristics as well as to compare our methodology to the results of \citet{2001MNRAS.326..901F} for the globular cluster 47 Tucanae. We then compare these results to those found using our MCMC fits in Section \ref{sec:Results}.

\subsection{Column Density of Pulsars}
\label{subsec:spectral_index}

We begin by calculating the average column density in an annulus around the CoG using the locations of our MSPs as well as the positions of bright X-ray sources as these have been shown to arise mostly from neutron stars in LMXBs \citep{1993SSRv...62..223L}. Thirty six X-ray sources were selected from \citet{2006ApJ...651.1098H}, using sources with ten or more counts between $0.5-6$ keV, as well as removing two X-ray sources coincident with known pulsars J1748$-$2446P and J1748$-$2446ad in order to avoid double counting \citep{2006ApJ...651.1098H}. We then measure $R_\perp$ for the X-ray sources and pulsars separately and calculate the cumulative distribution function (CDF) for the column density as a function of $R_\perp$.

\begin{figure}[t]
  \centering
  \includegraphics[width=\columnwidth]{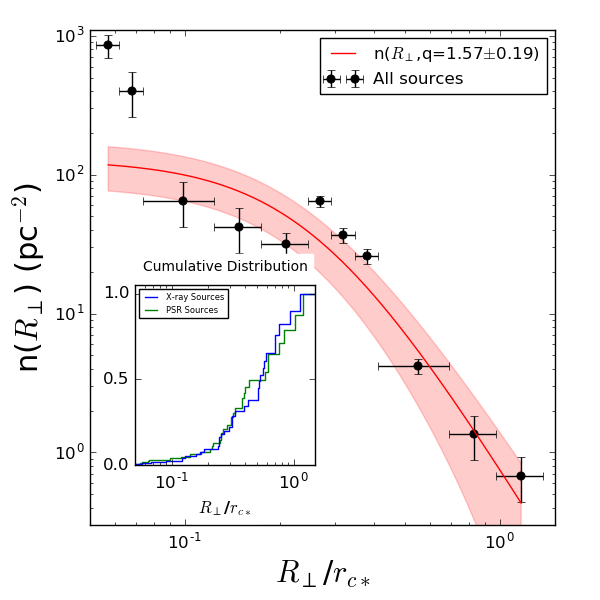}
  \caption
   { Column density of the combined MSP and LMXB populations as a function of $R_\perp/r_{c*}$ for Terzan 5. Over-plotted in red is Equation \ref{eq:king_cum} for a mass ratio of $\q$ and a dominant mass class core radius of r$_{c*}$=$\lanzonirc$. \citep{2010ApJ...717..653L}. The embedded plot shows the cumulative distributions of the pulsars and X-ray sources \citep{2006ApJ...651.1098H} separately. The p-value obtained from the two sample KS-test confirms that they are drawn from the same distribution. }
\label{fig:cum_dist}
\end{figure}

We verify that the X-ray sources and pulsars are drawn from the same parent density distribution by applying a two sample KS-test to the data. The embedded plot of Figure \ref{fig:cum_dist} shows the two distributions. We find a p-value of $\sim$0.99, meaning the pulsars and X-ray sources are likely drawn from the same parent distribution.

The observed column density profile for the neutron stars is given by Equation 1 from \citet{1995ApJ...439..191L}:
\begin{equation}
n(R_\perp)=n_0\left[1+\left(\frac{R_\perp}{r_{0}}\right)^2\right]^{\frac{\alpha}{2}}\,,
\label{eq:modified_power_law}
\end{equation}
where $r_0$ is a radial scale and is related to $r_c$. The spectral index for the neutron star population ($\alpha$) defines how centrally concentrated the pulsars are in the cluster.

We relate $r_0$ and $r_c$ by $r_c\equiv\left(2^{-2/\alpha}-1\right)^{1/2}r_0$. For $\alpha=-2$ this equation reduces to the single mass analytical King model \citep{1962AJ.....67..471K}.

For the multi-mass King model, the dominant mass class ($M_*$) in the cluster has a spectral index of $\alpha$=$-$2 and sets the potential for the GC most strongly. Different mass classes can be related to the dominant mass class according to the following relationships:
\begin{equation}
\begin{split}
r_{cX} &= \left(2^{-2/(1-3q)}-1\right)^{1/2}r_{c*} \\
\alpha_X &= 1-3q\,,
\end{split}
\label{eq:mass_relations_king}
\end{equation}
where we define q to be the mass ratio between a mass class $M_X$ and the dominant mass class $M_*$ ($q=M_X/M_*$). $r_{c*}$ is the core radius associated with $M_*$. \citep{2003ApJ...598..501H}. For the purposes of this work, we include both the isolated pulsars and the pulsar binaries in this calculation of $q$, as the neutron star mass is expected to have a wide distribution \citep{2016arXiv160302698O}, the companion mass is often much smaller than the pulsar mass, and the removal of the isolated pulsars from our sample does not significantly change our results.

We use the generalized form of Equation \ref{eq:modified_power_law} from \citet{2003ApJ...598..501H} to relate the pulsars to the dominant mass class:
\begin{equation}
n(r)=n_0\left[1+\left(\frac{R_\perp}{r_{c*}}\right)^2\right]^{\frac{1-3q}{2}}\,.
\label{eq:king_cum}
\end{equation}

\citet{2002ApJ...581..470G} found that the most visible stellar population in a cluster should have a spectral index of $\alpha=-2$. We therefore use the core radius ($\lanzonirc$) of bright main sequence turn off (MSTO) stars found by \citet{2010ApJ...717..653L} to approximate $r_{c*}$ and solve for the mass ratio of neutron stars and MSTO stars.

For the given value of $r_{c*}$=0.26 pc at a cluster distance of d=5.9 kpc, we perform a non-linear least squares fit for the expected number density of stars as a function of $n_0$ and $q$. We find $\q$, which gives a spectral index of $\alpha=\alphainit$. Figure \ref{fig:cum_dist} shows the best fit to Equation \ref{eq:king_cum} for both the pulsars and the X-ray sources combined.

This mass ratio agrees within the 1-$\sigma$ confidence intervals reported by \citet{2006ApJ...651.1098H} ($\qheinke$), which used only the LMXB population of Terzan 5 to calculate $q$. The turnoff mass of Terzan 5 is not well constrained, but using the results of \citep{2001ApJ...556..322B} of $M_*$=0.9 M$_\odot$, our mass ratio predicts an approximate pulsar or LMXB system mass of $\NSmass$. We will discuss the implications of our derived system mass in Section \ref{sec:discussion}.

\subsection{Core Density \& Core Radius}

\subsubsection{Procedure}
\label{subsubsec:core_param_initial}

In order to derive the core density and core radius of Terzan 5 using $R_\perp$ for each pulsar, we begin with Equation 3.5 from \citet{1993ASPC...50..141P}:
\begin{equation}
a_{\rm l,max} = \frac{2\pi G\rho_cr_c^2}{\sqrt{r_c^2+R_\perp^2}}\,,
\label{eqn:phin_amax}
\end{equation}
where the velocity dispersion $\sigma$ from Equation 3.5 \citep{1993ASPC...50..141P} was replaced. This form of the acceleration equation is accurate to within 10\% for $r<2r_c$ and to within 50\% for intermediate radii and is useful when line of sight positions for each source are unknown.

From Equation 3.12 from \citet{1993ASPC...50..141P} we find the probability of having a measured $a_{\rm l}$ less than $a_{\rm l,max}(R_{\perp})$ for our measured value of $\alpha$:
\begin{equation}
\begin{split}
&p(a_l|a_{\rm l,max}(R_{\perp}))d\left|a_l\right| = P(\alpha)\left(\frac{a^{\alpha-2}}{\sqrt{1-a^2}}\right) \\
&\times\left[\left(1-\sqrt{1-a^2}\right)^{1-\alpha/2}+\left(1+\sqrt{1-a^2}\right)^{1-\alpha/2}\right]da\,,
\end{split}
\label{eq:pdf}
\end{equation}
where $a=a_{\rm l}$/$a_{\rm l,max}(R_{\perp})$ and $P(\alpha)$ is a normalization constant given by:
\begin{equation}
P(\alpha)=\frac{\Gamma(\alpha/2)2^{(2-\alpha)/2}}{\sqrt{\pi}\Gamma([\alpha-1]/2)}\,,
\end{equation}
where $\Gamma$ is the gamma function.

\begin{figure}[t]
  \centering
  \includegraphics[width=\columnwidth]{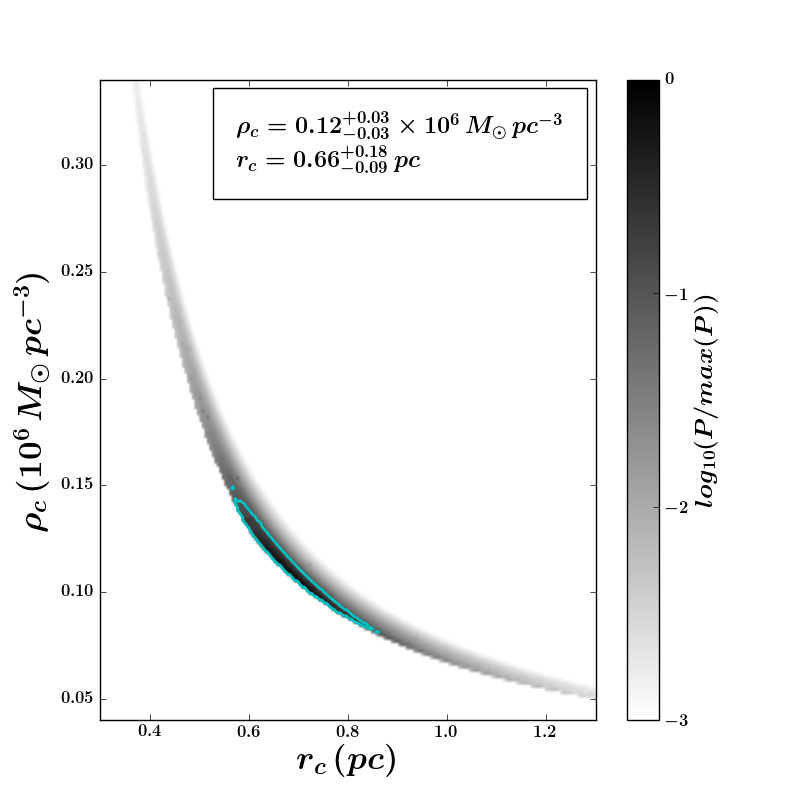}
  \caption
  {The log likelihood for all possible combinations of model parameters $\rho_c$ and $r_c$ for 47 Tucanae. The contour shows the 1-$\sigma$ confidence region. We take the median of the marginalized parameters as our best fit and the extent out from the median to the 1-$\sigma$ contour region as the errors. When using only projected positions of each pulsar, there is strong covariance between $\rho_c$ and $r_c$. }
\label{fig:loglike_47tuc}
\end{figure}

Using the measured value of $\dot{P}_b$ when available and solving Equation \ref{eq:pdot_meas_terms} assuming the median expected magnetic field strength of our isolated pulsars (Section \ref{subsubsec:intrinsic_spindown}), we obtain a measurement of the cluster-only acceleration to use with Equation \ref{eq:pdf} for all of our pulsars.

We find the probability of a given set of cluster parameters producing each pulsar's measured acceleration by multiplying the result of Equation \ref{eq:pdf} with the measurement error associated with $a_{\rm l}$ from the timing noise and the removal of $a_g$, $a_s$, and {\scriptsize $\left(\dot{P}/P\right)_{\rm int}$}. We then convert this probability into a log likelihood and sum up the total log likelihood for all of the pulsars for a given set of model parameters. 

Errors in our cluster parameters are calculated by finding the contour region about the maximum likelihood that contains 68\% of the total probability in our simulation grid. We report the median of the marginalized distribution of each parameter as the best fit to the cluster potential, with the extent of the 1-$\sigma$ confidence interval out from the median as the errors in each parameter.

\subsubsection{47 Tucanae}
\label{sec:projected_47tuc}

\begin{figure}[t]
  \centering
  \includegraphics[width=\columnwidth]{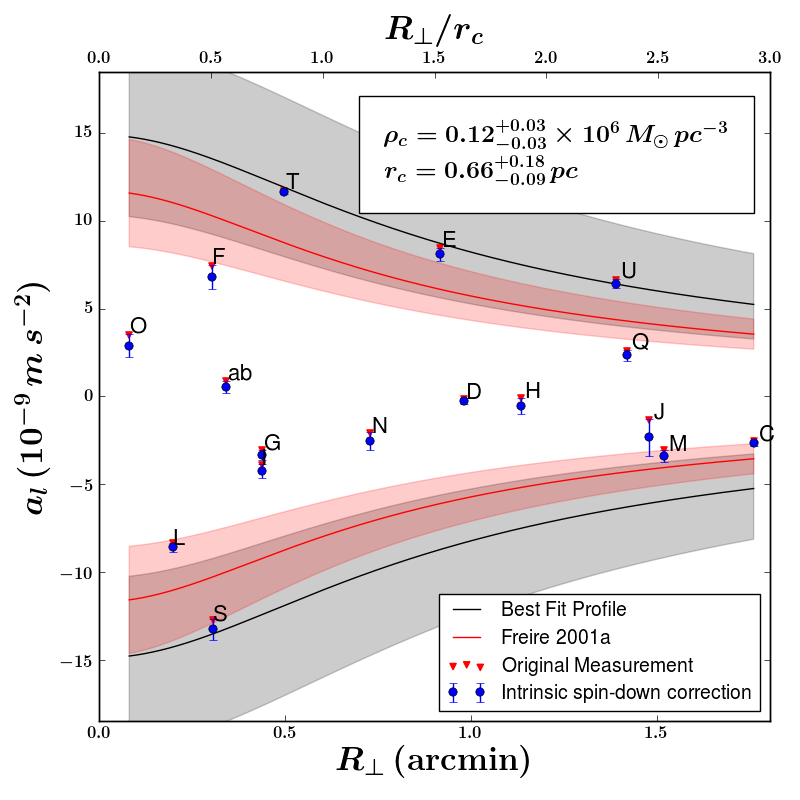}
  \caption
  { The best acceleration profile for 47 Tucanae using Equation \ref{eqn:phin_amax}. Plotted in red are the measured $a_{\rm l}$ values and in blue are the values of $a_{\rm c}$ for the pulsars where we modeled $P_{\rm int}$ using the ATNF catalog. Shaded regions show the 1-$\sigma$ confidence interval for our fits. Over-plotted in red is the fit used previously by \citet{2001MNRAS.326..901F}. When using only the projected positions for each pulsar and no additional information from optical measurements, there is strong covariance between the core density and core radius. }
\label{fig:accel_prof_47tuc}
\end{figure}

Before analyzing the timing data of Terzan 5, we compare how well our model reproduces results found previously using the 25 known pulsars in 47 Tucanae. We use the timing properties derived by \citet{2003MNRAS.340.1359F} and \citet{2016MNRAS.459L..26P} for each pulsar referenced to a CoG located at $\CoGRATuc$ and $\CoGDECTuc$ \citep{2006AJ....131.1163S}.

\citet{2005ApJ...625..796H} found that for 47 Tucanae $q=1.63\pm0.11$, giving a spectral index of $\alpha=-3.9\pm0.33$. For the timing solutions provided by \citet{2003MNRAS.340.1359F}, Equation \ref{eq:pdf} gives the log-likelihood plane defined by our trial values of $\rho_c$ and $r_c$ for this cluster. Figure \ref{fig:loglike_47tuc} shows the results of our analysis.

Without knowing the true line of sight position for each pulsar, large covariances between $\rho_c$ and $r_c$ can arise. Also, if a system does not have any pulsars at small values of $R_\perp$ with large values of $a_{\rm l}$ near $a_{\rm l,max}(R_{\perp})$, we cannot limit the allowed parameter space in terms of the central density very strongly.

Our analysis finds a peak likelihood at $\tucrho$ and $\tucrc$. \citet{2001MNRAS.326..901F} found in a similar analysis a core radius of $r_c=0.6\pm0.04$ pc, though in their work they assume a fixed value of $\rho_c=$10$^5$ M$_\odot$ pc$^{-3}$ as measured by \citet{2001MNRAS.326..901F}. Comparing our results to the literature, we find good agreement for both cluster core parameters.

Figure \ref{fig:accel_prof_47tuc} shows the best fit acceleration profile along with the measured pulsar accelerations. We have also plotted the best fit acceleration profile used by \citet{2001MNRAS.326..901F} in red. We find good agreement with their results to within the 1-$\sigma$ confidence interval, which are plotted as the shaded regions.

\subsubsection{Terzan 5 Results}
\label{subsubsec:initial_results}

\begin{figure}[t]
  \centering
  \includegraphics[width=\columnwidth]{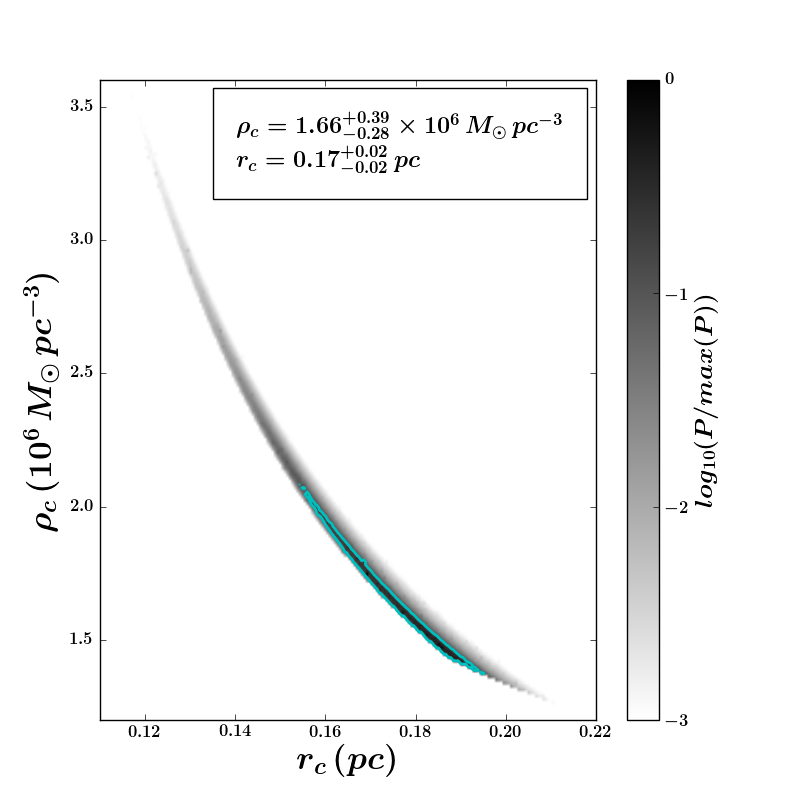}
  \caption
  {The log likelihood for all possible combinations of model parameters $\rho_c$ and $r_c$ for Terzan 5. The contour shows the 1-$\sigma$ confidence region. We take the median of the marginalized parameters as our best fit and the extent out from the median to the 1-$\sigma$ contour region as the errors. When using only the projected positions for each pulsar and no additional information from optical measurements, there is strong covariance between the core density and core radius. }
\label{fig:loglike}
\end{figure}

We show the results of our search over the likelihood space for the core density and core radius for Terzan 5 in Figure \ref{fig:loglike} with the spectral index found in Section \ref{subsec:spectral_index}.

The maximum likelihood was found at $\rho_c$=$\dencinit\times$10$^6$ M$_\odot$ pc$^{-3}$ and $r_c=\rcinit$ pc, which corresponds to an angular size of $\rcinitarcsec$ arcseconds at the cluster distance of 5.9 kpc. Table \ref{table:cluster_params} compares our results against previously quoted values from the literature, finding good agreement for each parameter.

\begin{figure}[t]
  \centering
  \includegraphics[width=\columnwidth]{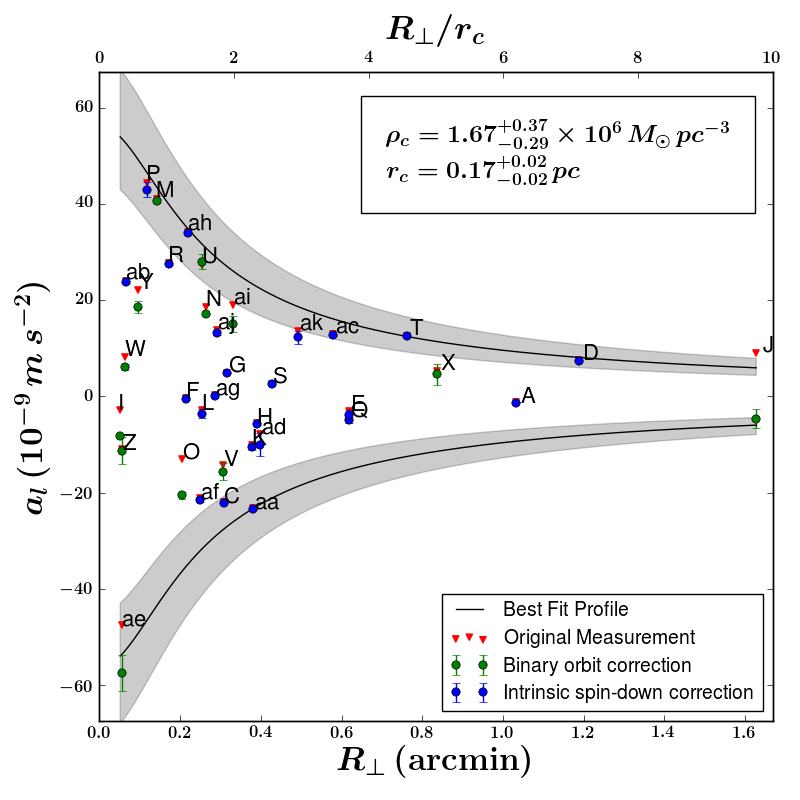}
  \caption
  { The best acceleration profile for Terzan 5 using Equation \ref{eqn:phin_amax}. Plotted in red are the measured $a_{\rm l}$ values, in green are the values of $a_c$ for pulsars with a measured $\dot{P}_b$, and in blue are the values of $a_c$ for the remaining pulsars where we modeled $P_{\rm int}$ using the ATNF catalog. }
\label{fig:accel_prof}
\end{figure}

The acceleration profile associated with the maximum likelihood is shown in Figure \ref{fig:accel_prof}. We can see that there are only a few pulsars near the maximum acceleration that provide the strongest constraints for this method of analyzing the data. Using a properly de-projected position for each pulsar will allow us to gain more insight into the cluster dynamics.

\subsection{Line of Sight Positions}
\label{sec:init_los}

We can estimate the line of sight positions of each pulsar relative of the plane O of the cluster by solving Equation \ref{eq:king_amax} for the line of sight positions that give our measured accelerations at a given $R_\perp$.

Table \ref{table:initial_pos} shows the initial best fits for the line of sight position of each pulsar using the core density, core radius, and spectral index derived in Sections \ref{subsec:spectral_index} and \ref{subsubsec:initial_results}. We also include the relative probability of finding a pulsar at one of the two line of sight positions according to the prescription of Appendix D from \citet{1993PhDT.........2A}:
\begin{equation}
p(l|R_\perp,r_c,\alpha)dl=\frac{n_pdl}{\int_{-\infty}^\infty n_pdl}\,,
\label{eqn:pdf_los}
\end{equation}
where $n_p$ is the pulsar number density given by:
\begin{equation}
n_p\propto\left(r_c^2+R_\perp^2+l^2\right)^{\alpha/2}\,.
\end{equation}

This provides us with two solutions for the line of sight position, where each peak has roughly equal probabilities to within a factor of a few. Using our measurements of jerks, we can remove some of this degeneracy as the radial offset from the CoG factors heavily into the expected jerk felt by each pulsar. We discuss the resulting likelihood for the position of each pulsar in Sections \ref{subsubsec:pos_like} and \ref{subsubsec:jerk_like} and introduce how we overcome the difficulty of modeling bimodal solutions in Section \ref{sec:PTSampling}.

\section{MCMC Analysis}
\label{sec:MCMC}

In order to better derive the cluster parameters for Terzan 5 we used the sampling package {\tt emcee} \citep{2013PASP..125..306F} to de-project our pulsar positions and obtain a fully 3-dimensional model for the cluster's density profile. We then directly solve for the acceleration produced by the mass interior to each pulsar. De-projecting each pulsar individually and comparing its measured acceleration to the model acceleration allows us to reduce the amount of covariance in our cluster parameters. To begin, we discuss the individual likelihood functions used and their priors before introducing the actual simulations and their results.

\subsection{Likelihoods}

From our pulsar timing data, we consider only the spin period, the spin period derivative, the second-order spin period derivative, the orbital period, and the orbital period derivative when defining the likelihood. The total probability of measuring a given on-sky position, acceleration ($a_r$), and a jerk of a pulsar ($\dot{a}_r$, if measured) is given by
\begin{equation}
\begin{split}
p(R_\perp,a|\rho_c,r_c,\alpha)& = p(l|R_\perp,r_c,\alpha)\times p(R_\perp|r_c,\alpha)\times \\
&p(a_r|\rho_c,r_c,\alpha)\times p(\dot{a}_r|\rho_c,r_c,\alpha)\,,
\end{split}
\label{eq:total_prob}
\end{equation}
where the first two terms are the probability associated with the position of the pulsar in the potential, and the final two terms give the probability of measuring a given acceleration or jerk respectively.

We defined the log-likelihood ($\mathcal{L}$) as follows:
\begin{equation}
\mathcal{L} = \mathcal{L}_{\rm pos} + \mathcal{L}_{\rm accel} + \mathcal{L}_{\rm jerk}\,,
\label{eq:total_loglike}
\end{equation}
where $\mathcal{L}$ is the total log-likelihood, $\mathcal{L}_{\rm pos}$ is the likelihood associated with the first two terms of Equation \ref{eq:total_prob}, $\mathcal{L}_{\rm accel}$ is the log likelihood from our acceleration measurements, and $\mathcal{L}_{\rm jerk}$ is the log likelihood from our jerk measurements.

\subsection{Positional Likelihood}
\label{subsubsec:pos_like}

In order to calculate the likelihood of finding a pulsar at its observed on-sky location $R_\perp$ and at a given line of sight position $l$, we decomposed the likelihood for position into two components:
\begin{equation}
\mathcal{L}_{\rm pos} = \mathcal{L}_l + \mathcal{L}_{\rm R_\perp}\,,
\label{eq:pos_loglike}
\end{equation}
where $\mathcal{L}_l$ is the likelihood for the line of sight position given by Equation \ref{eqn:pdf_los}.

$\mathcal{L}_{\rm R_\perp}$ is the likelihood for the on-sky position of each pulsar and is given by:
\begin{equation}
p(R_\perp|r_c,\alpha)dR_\perp=\frac{n_pdR_\perp}{\int_{-\infty}^\infty n_pdR_\perp}\,,
\label{eqn:pdf_rperp}
\end{equation}
where $n_p$ is the number density of pulsars and is readily defined by the on-sky form of the King model \citep{1962AJ.....67..471K} which is proportional to
\begin{equation}
n_p\propto\left(1+\left(\frac{R_\perp}{r_c}\right)^2\right)^{\alpha/2}\,.
\end{equation}

If fitting for the offset in the CoG of the cluster, we measure a new spherical angular difference for each pulsar from the CoG to obtain a new value of R$_\perp$ before applying this likelihood function. In Section \ref{subsec:priors} we discuss the priors used on the allowed parameter space for the offset in the CoG position.

Converting Equation \ref{eqn:pdf_los} into a likelihood, we find
\begin{equation}
\mathcal{L}_l \propto \sum_i \ln \left[(r_c^2+R_{\perp,i}^2+l_i^2)^{\alpha/2}\right]\,,
\end{equation}
for the line of sight probability, where i is the summation over each pulsar in the system.

For the on-sky positional probability, the likelihood is given by taking the log of Equation \ref{eqn:pdf_rperp}:
\begin{equation}
\mathcal{L}_{R_\perp} \propto \sum_i \ln\left[ \left(1+\left(\frac{R_{\perp,i}}{r_c}\right)^2\right)^{\alpha/2}\right]\,.
\end{equation}


\subsubsection{Acceleration Likelihood}
\label{subsubsec:Accel_like}

Using the results of Section \ref{subsec:sim_accel} we identified that only the mean-field component of the acceleration is important for our analysis. As such, we further decomposed the acceleration likelihood into two components:
\begin{equation}
\mathcal{L}_{\rm accel} = \mathcal{L}_{\rm binary} + \mathcal{L}_{\rm spindown}\,,
\label{eq:accel_loglike}
\end{equation}
where $\mathcal{L}_{\rm binary}$ is the log likelihood from the binary pulsars with measured values of $\dot{P}_b$ and $\mathcal{L}_{\rm spindown}$ is the log likelihood from the remaining pulsars.

Equation \ref{eq:king_amax} gives the model acceleration for the given cluster parameters and pulsar position $r$ in the cluster. The position $r$ in Equation \ref{eq:king_amax} is found by applying the Pythagorean theorem to $R_\perp$ and the line of sight estimate $l$ from MCMC sampling.

Using a normal distribution about the predicted acceleration $\hat{a}(r|\theta)$ for a given set of cluster parameters $\theta$ we derive the likelihood for the binary pulsars:
\begin{equation}
\mathcal{L}_{\rm binary} \propto \sum_i \frac{1}{2\epsilon_{\rm i}^2}\left(a_{\rm l,i}-\hat{a}(r|\theta)\right)^2\,,
\label{eq:binary_log_eqn}
\end{equation}
where $i$ is the summation over each binary pulsar with a measured $\dot{P}_b$ in the cluster, $\epsilon$ is the error in each measured acceleration, and $\theta$ are the current cluster parameters to be tested.

For the remaining pulsars, Equation \ref{eq:king_amax} gives the model acceleration for the given cluster parameters and pulsar position $r$ in the cluster and Equation \ref{eq:pdot_meas_terms} gives the measured cluster acceleration for those systems without a measured $\dot{P}_b$.

While the terms due to the Galactic acceleration and Shklovskii effect can be removed from the measured acceleration, the intrinsic spin-down due to magnetic braking is not necessarily known. Subtracting the model acceleration from the measured accelerations must then leave a residual component due to this effect, which is the dominant source of error in our measurements of the cluster acceleration.

This intrinsic spin-down component can be modeled using the magnetic field strength given by Equation \ref{eq:spindown}, which follows a log-normal distribution that we can use to calculate a likelihood estimate according to the pulsar catalogue parameters shown in Figure \ref{fig:atnf_bfield_dist}.

The log-likelihood of this log-normal distribution of magnetic field strengths is given by:
\small
\begin{equation}
\mathcal{L}_{\rm isolated} \propto \sum_i \left[\frac{1}{2(\sigma_B)^2}\left(\log{B_{8}}-\mu_B\right)^2+\log{B_{8}}\right]\,,
\label{eq:iso_log_eqn}
\end{equation}
\normalsize
where $B_8$ is defined to be $B/(10^8 \,{\rm Gauss})$. 

\subsubsection{Jerk Likelihood}
\label{subsubsec:jerk_like}

In Section \ref{subsec:sim_jerk} we showed that by removing the mean-field component of the jerk from the total jerk felt by a pulsar, the residuals follow a Lorentzian given by Equation \ref{eqn:jerk_pdf}. The log likelihood for the jerk is then defined to be:
\small
\begin{equation}
\mathcal{L}_{\rm jerk} = \sum_i \log\left(\frac{\dot{a}_0}{\pi}\frac{1}{\left[\left(\dot{a}_{l,i}-\dot{a}_{{\rm mf},i}\right)^2+\dot{a}_0^2\right]}\right)\,.
\label{eq:iso_log_eqn}
\end{equation}
\normalsize

Using this likelihood we may have additional fitting power for each pulsar's line of sight position and therefore the cluster parameters. Using only accelerations, two different positions along our line of sight can produce the same value (Figure \ref{fig:layout}), whereas using a measured jerk allows for additional information about the pulsar's position, potentially breaking this degeneracy.

\subsubsection{Central Black Hole Models}

Following the prescription of Section \ref{sec:theory_bh}, we introduced an additional fitting parameter for a point-like central mass in the system, allowing the density profile to vary accordingly. The likelihood for this was calculated using Equation \ref{eq:accel_loglike}. The model acceleration for both $\mathcal{L}_{\rm binary}$ and $\mathcal{L}_{\rm spindown}$ were calculated using Equation \ref{eqn:model_accel_bh}.

\subsection{Priors}
\label{subsec:priors}

Flat priors were used for the range of cluster parameters, pulsar positions, and black hole mass estimates. For all but the black hole mass estimates these priors were flat in linear spacing, whereas fits for the central black hole mass were performed with log-uniform flat spacing. The log-uniform prior was used to properly sample over the many orders of magnitude possible for a central black hole mass.

For the three main cluster parameters, the flat priors ranged from 10$^{4}$ $\leq\rho_c\leq$ 10$^{7}$ M$_\odot$ pc$^{-3}$, 0 $\leq r_c\leq$ 2 pc, and $-8\leq\alpha\leq-2$. The range in our spectral index was chosen to slightly expand upon the results of \citet{1993ASPC...50..141P}, which found pulsars should have values of $-6<\alpha<-2$.

In the case that the cluster's CoG was fit for, a normal prior centered on a zero offset with a FWHM of $1'$ was used for the allowed variations in right ascension and declination.

\subsection{MCMC Parameters}
\label{subsec:MCMC_params}

Our simulations were initialized using $\nwalkeraccel$ walkers. When fitting for the cluster parameters of Terzan 5 using only the acceleration measurements, we fit for $\ndimaccel$ parameters, which include the core density, core radius, spectral index, and pulsar line of sight positions. 

For acceleration plus jerk fittings, we used the same parameters as the acceleration-only fits, while including a line of sight velocity estimate for each pulsar with a measured $\dot{a}_{\rm l}$. This gives a total number of $\ndimjerk$ parameters to fit for Terzan 5.

If performing a fit for the CoG, we used the same parameters as the acceleration-only fit, with two additional parameters that identify the offset in right ascension and declination. This gives a fit with 39 parameters for Terzan 5.

We also performed a simulation which looked for the presence of a central black hole in the system using the measured accelerations and jerks. This fit used all the previously described parameters, while including one additional parameter for the black hole mass, giving a total of $\ndimbh$ parameters for Terzan 5.

The initial parameter estimates were found using the results of Section \ref{sec:initial_fits} in order to reduce the required time for burn in. For the line of sight position of each pulsar, we assigned a line of sight position normally distributed around one of the two possible line of sight positions for each pulsar shown in Table \ref{table:initial_pos}. The number of walkers centered about a given line of sight position was selected according to the relative probability between $l_1$ and $l_2$.

\subsection{Parallel Tempering}
\label{sec:PTSampling}

We addressed the issue of multi-modality in the line of sight position by adding parallel tempering to our simulations. When modeling parameters with strong multi-modality, the popular method of a single MCMC chain stochastically exploring the parameter space is an ill-suited approach as a single chain is unable to explore separate peaks in the log-likelihood parameter space.

If two solutions are separated by many standard deviations, a chain may end up stuck in a local over-density of probability. Using a parallel tempered sampler however, we can allow chains to more fully explore the parameter space \citep{1992EL.....19..451M}. Using the package {\tt PTSampler} provided by \citet{2013PASP..125..306F}, we include an additional dimension to our simulation, which varies according to a temperature T. By applying an exponential factor to the likelihood function $\mathcal{L}^{1/T}$, the likelihood function becomes shallower and broader for higher temperatures. 

The higher-temperature chains are then allowed to fully explore the parameter space, and by exchanging positions with colder chains, the chain with the true likelihood function (T=1) is eventually able to fully explore the parameter space. 

For our models, we used $\ntemp$ different temperatures in order to fully explore the parameter space.

\section{MCMC Results for 47 Tucanae}
\label{sec:mcmc_47tuc}

\begin{figure}[t]
  \centering
  \includegraphics[width=\columnwidth]{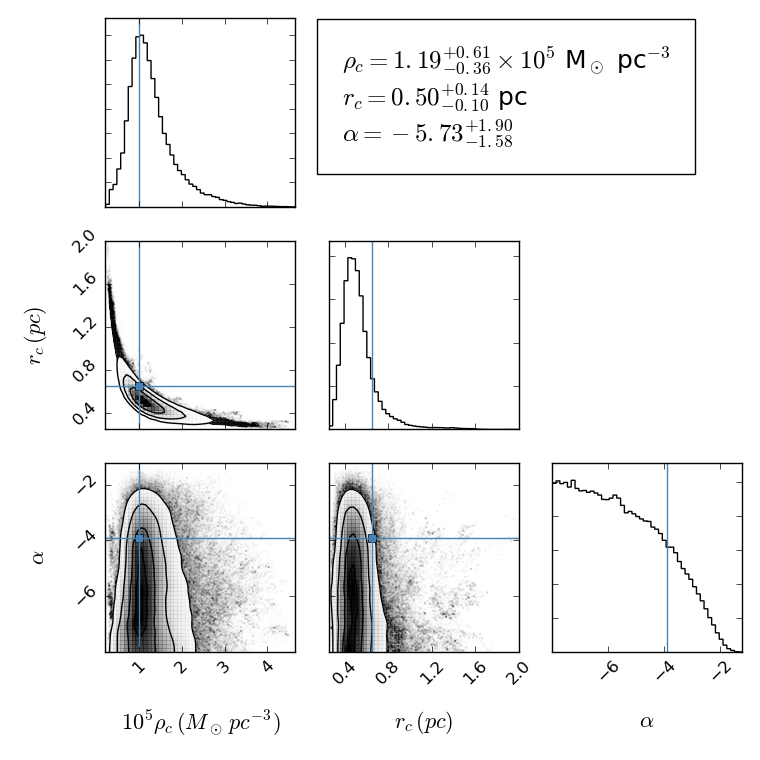}
  \caption
  { Corner plot showing the marginalized probabilities of finding a given cluster parameter for the cluster 47 Tucanae, along with the covariances between different parameters. These results use only the pulsar accelerations. Blue lines show previously derived estimates of each parameter from \citet{2001MNRAS.326..901F}. }
\label{fig:tuc47_corner}
\end{figure}

In order to verify that our MCMC method can produce valid results for the Terzan 5 system, we first performed the analysis on the measured accelerations within 47 Tucanae. The results of our simulation are shown in a corner plot \citep{dan_foreman_mackey_2014_11020} in Figure \ref{fig:tuc47_corner}. We find $\rho_c$=$\denctuc\times$10$^5$ M$_\odot$, $r_c$=$\rctuc$ pc, and $\alpha$=$\alphatuc$. These results agree to within the 1-$\sigma$ confidence intervals of those found by \citet{2001MNRAS.326..901F} (Shown in blue in Figure \ref{fig:tuc47_corner}), as well as with our initial estimates using the projected positions of each pulsar in Section \ref{sec:projected_47tuc}, for $\rho_c$ and $r_c$.

In the case of $\alpha$, we find our results to be largely unconstrained. Our flat prior from $-8<\alpha<-2$ is only constrained at the lowest values of $\alpha$. We argue that the reason for this is likely due to the lack of pulsars with large values of $a_{\rm l}$ at small radii. Without more pulsars, we are unable to constrain the fits for the spectral index strongly. With the recent discovery of two new pulsars in this cluster however, it is promising that future studies will help to constrain these parameters further \citep{2016MNRAS.459L..26P}. In Terzan 5 however, the large number of pulsars should provide a stronger contraint on $\alpha$ due to the combined positional probabilities.

\section{MCMC Results for Terzan 5}
\label{sec:Results}

In this section we present the results of our MCMC analysis using only the measured accelerations to begin with, followed by the simultaneous analysis of the pulsar accelerations and jerks (if measured). We then compare these results to one another, as well as with the results from traditional pulsar timing techniques. Lastly, we discuss the results from our fits to a central black hole in Terzan 5 and discuss the implications of having such a black hole in the system.

All of our results are summarized in Table \ref{table:cluster_params}.

\subsection{Acceleration Only}

\begin{figure}[t]
  \centering
  \includegraphics[width=\columnwidth]{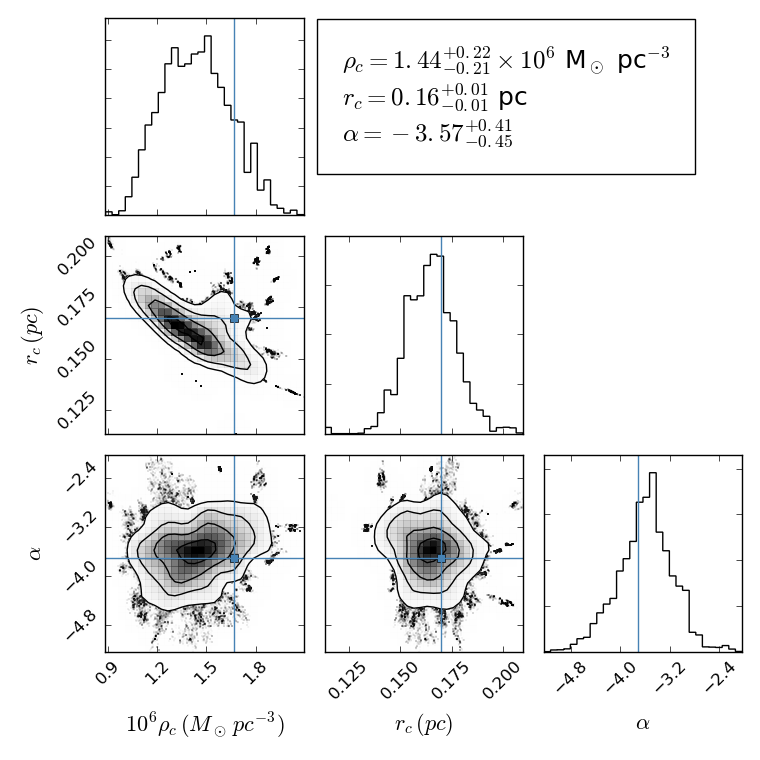}
  \caption
  { Corner plot showing the marginalized probabilities of finding a given cluster parameter for the cluster Terzan 5, along with the covariances between different parameters. These results use only the pulsar accelerations.  Blue lines show previously derived estimates of each parameter from Section \ref{subsubsec:initial_results} using the projected position of each pulsar. }
\label{fig:accel_only_corner}
\end{figure}

The cluster parameters found using the measured accelerations and the likelihood functions of Section \ref{subsubsec:Accel_like} are shown in Table \ref{table:cluster_params}. We also compared our results to previously cited values in the literature as well as the values found using only the projected position of each pulsar and its acceleration. We also include a corner plot \citep{dan_foreman_mackey_2014_11020} in Figure \ref{fig:accel_only_corner} to show the marginalized PDFs of the three main cluster parameters as well as their covariances. The 1-$\sigma$ confidence intervals were integrated out from the median of the marginalized distribution for each position. 

Table \ref{table:results} shows the line of sight position of each pulsar, sorted by the pulsar's measured value of $R_\perp$, after each walker completed its analysis of the parameter space. We also include the relative number of walkers in the $l_1$ and $l_2$ peaks as the ratio of probabilities in these two solutions (P$_{\rm ratio}$).

\begin{figure}[t]
  \centering
  \includegraphics[width=\columnwidth]{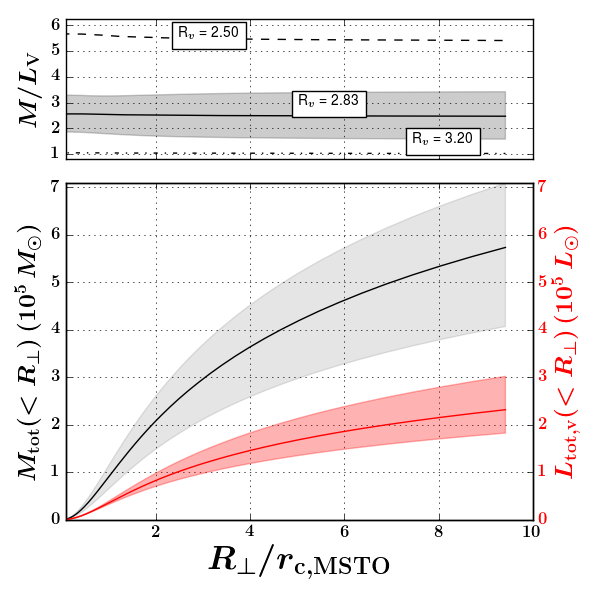}
  \caption
  { Mass and V-band luminosity profile as a function of $R_\perp$ using only the acceleration measurements shown in black on the left hand axis and in red on the right hand axis, respectively. Each profile has been expressed in terms of the dominant mass class within the cluster. We find a mass to light ratio of M/L$\simeq$2.8 in the core of the cluster. The top panel shows the systematic offsets in our median mass to light ratio using R$_v$ (and not the extinction A$_v$) due to the strong dependence on the color excess E(B-V) depending on the line of sight through the cluster. }
\label{fig:ML_accel}
\end{figure}

Our MCMC analysis finds $\rho_c$=($\dencaccel$)$\times$10$^6$ M$_\odot$ pc$^{-3}$, which is slightly lower, but in agreement with, our initial estimates of $\rho_c$=($\dencinit$)$\times$10$^6$ M$\odot$ pc$^{-3}$. These results are also consistent within the 1-$\sigma$ confidence interval given by \citet{2010ApJ...717..653L}. We find that our errors using the MCMC sampler are signficantly improved over these previous methods of study.

Our measurement of $\alpha$=$\alphaaccel$ is consistent with the one found using traditional projected techniques of $\alpha$=$\alphainit$ though with an improved error estimate. Our spectral index from MCMC fitting gives a mass ratio of $q$=$\qaccel$ which agrees with \citet{2006ApJ...651.1098H}. For a MSTO mass of 0.9 M$_\odot$, this gives a median pulsar system mass of $\sim$1.3 M$_\odot$.

Our core radius of $r_c$=$\rcaccel$ pc using the MCMC analysis and acceleration only measurements is also consistent with the initial estimate $r_c$=$\rcinit$ pc. We compare our results to those of \citet{2010ApJ...717..653L} and \citet{2013ApJ...774..151M} where the core radius of MSTO stars were measured using Equation \ref{eq:mass_relations_king} and the measured mass ratio. For a mass ratio of $\qaccel$, we find a MSTO star core radius of $r_{c,*}$=0.23$\pm$0.02 pc. This agrees with \citet{2013ApJ...774..151M}, who found a core radius of 0.22$\pm$0.01 pc.

\begin{figure}[t]
  \centering
  \includegraphics[width=\columnwidth]{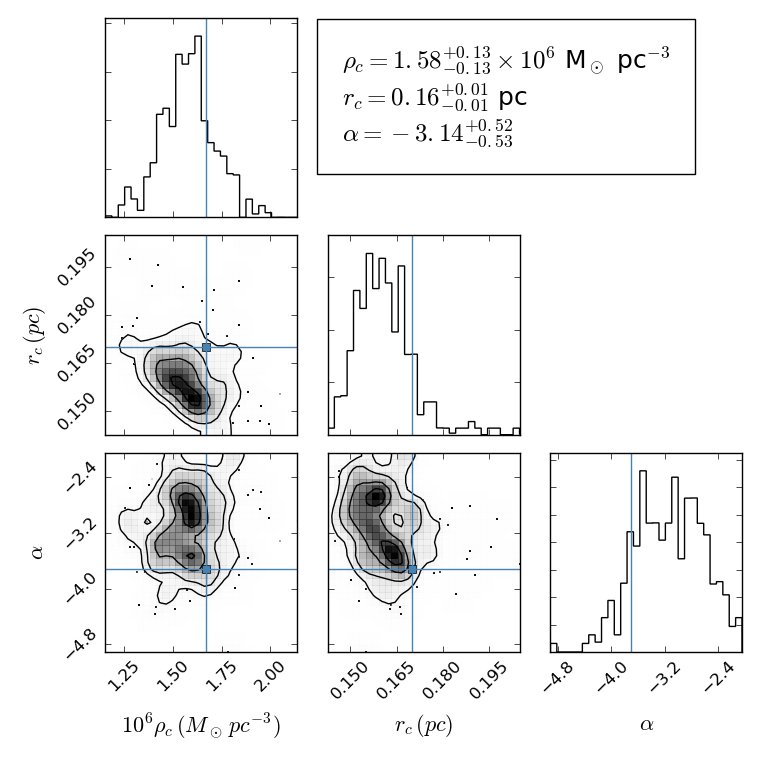}
  \caption
  { Corner plot showing the marginalized probabilities of finding a given cluster parameter for Terzan 5, along with the covariances between different parameters. These results use the pulsar accelerations as well as their jerks (when available). Blue lines show previously derived estimates of each parameter from Section \ref{subsubsec:initial_results} using the projected position of each pulsar. }
\label{fig:jerk_corner}
\end{figure}

In order to calculate the total mass of the cluster as a function of $R_\perp$, we expressed our results in terms of the classical single-mass King model. As we have solved for a mass class other than the dominant one, we scale our core radius to the expected dominant core radius according to Equation \ref{eq:mass_relations_king}. We also use the expected spectral index of the dominant mass class of $\alpha=-2$. We then integrate up all the mass along within a given on-sky position $R_\perp$. 

Figure \ref{fig:ML_accel} shows the mass within the cluster as a function of $R_\perp$. We find that the integrated mass along our line of sight within a cylindrical radius of 1 parsec to be M($R_\perp<$1pc)$\sim$3.5$\times$10$^5$ M$_\odot$.

We have also calculated the observed V-band luminosity of Terzan 5 at matching radii using the measured surface brightness in the V-band from \citet{2010ApJ...717..653L}. We use the results of \citet{2012ApJ...755L..32M} to find the color excess E(B-V) as a function of the radial distance from the CoG for an assumed extinction coefficient R$_v$=2.83 to convert the measured surface brightness into a luminosity measurement. Errors were found by taking the minimum and maximum value of E(B-V) in a radial bin out from the CoG.

We find a central mass to light ratio of M/L$_v\simeq$2.6$_{-0.7}^{+0.7}$. It should be noted however that the exact value of R$_v$ can drastically change our results for the mass to light ratio as shown in the top plot of Figure \ref{fig:ML_accel}. Without a more constraining measurement for the extinction towards this region of the galaxy, our results are dominated by the errors in the photometry. We compare our results to those of \citet{2009AJ....138..547S}, who find a M/L$_v\simeq$1.75$\pm$0.1 for metal-rich clusters. Given the additional errors in R$_v$, we consider this to be an acceptable agreement for the metal rich Terzan 5.

The line of sight positions for each pulsar are also shown in Table \ref{table:results}. We find that most of the pulsars lie close to the plane of the sky O as defined by the relative probability between the $l_1$ and $l_2$ positions. By including the jerks, we will investigate whether we can even better constrain the position of each pulsar along our line of sight.

\subsection{Accelerations and Jerks}

Starting with the same initial conditions as the acceleration only measurements (Table \ref{table:initial_pos}), a second simulation was run including jerk measurements according to the prescription of Section \ref{subsubsec:jerk_like}. Of the $\Npsr$ pulsars, 32 have measured jerks. Pulsars without measured jerks were analyzed using only their accelerations and provided no weight to the likelihood for jerk measurements. Our results for this simulation are shown in Figure \ref{fig:jerk_corner} and Tables \ref{table:cluster_params} and \ref{table:results}.

\begin{figure}[t]
  \centering
  \includegraphics[width=\columnwidth]{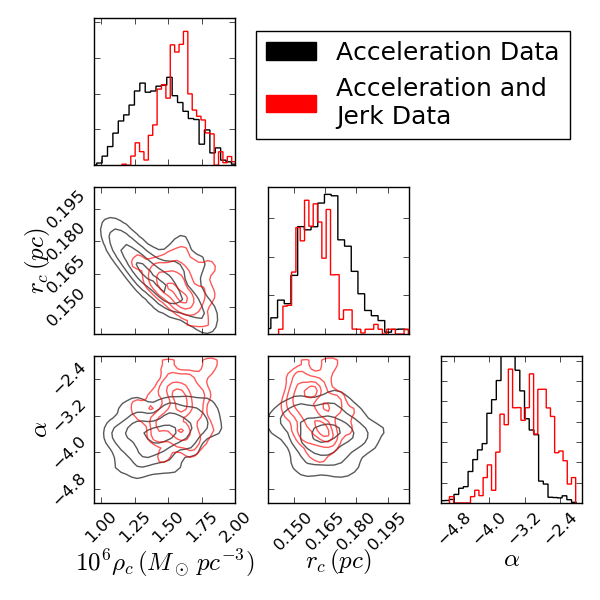}
  \caption
  { Posterior distributions of the core density, core radius, and the spectral index for simulations using only accelerations in black and accelerations plus jerks in red. We do not show individual data points for the acceleration plus jerk fits for clarity. }
\label{fig:corner_stack}
\end{figure}

Using the jerks in addition to the accelerations, we found $\rho_c$=($\dencjerk$)$\times$10$^6$ M$\odot$ pc$^{-3}$, $r_c$=$\rcjerk$ pc, and $\alpha$=$\alphajerk$. All of these values are in good agreement with those found using only the measured pulsar accelerations and are compared to the literature in Table \ref{table:cluster_params} . Figure \ref{fig:corner_stack} shows an overlay of the two simulations, where we see slightly more constraining fits when analyzed using pulsar jerks.

We also find a minor bimodality arising in our fits of the specral index $\alpha$ when jerk fitting is turned on. This is likely due to the fact that the pulsar jerks provide additional line of sight information and may imply that some of our pulsars must lie at greater distances from the core in order to have a given value of jerk that goes beyond the line of sight probability given by Equation \ref{eqn:pdf_los}.

This implies that future studies of the GC potential will benefit from long-term timing of pulsars, as the jerks provide additional information about the overall dynamics of the system. Given the relationship between the dominant mass class and the neutron star systems (Equation \ref{eq:mass_relations_king}), additional timing of pulsars and the MSTO stars in GCs may provide additional constraints on the neutron star mass distribution.

\begin{figure}[t]
  \centering
  \includegraphics[width=\columnwidth]{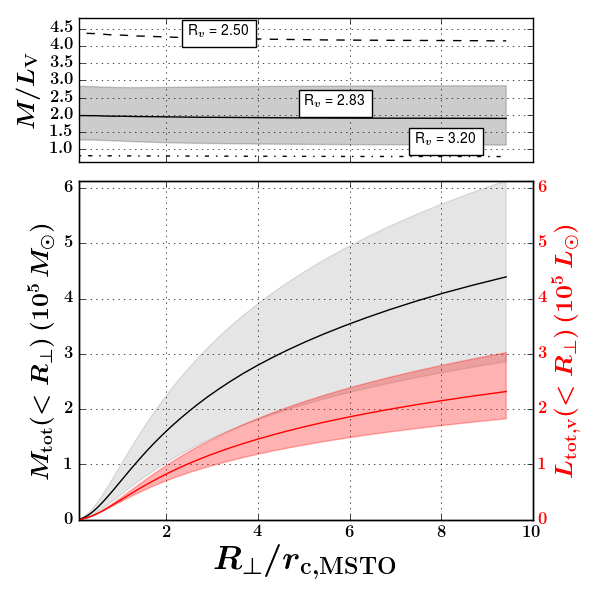}
  \caption
  { Mass and V-band luminosity profile as a function of $R_\perp$ using the acceleration and jerk measurements shown in black on the left hand axis and in red on the right hand axis, respectively. Each profile has been expressed in terms of the dominant mass class within the cluster. We find a mass to light ratio of M/L$\simeq$2.6 in the core of the cluster.  The top panel shows the systematic offsets in our median mass to light ratio using R$_v$ (and not the extinction A$_v$) due to the strong dependence on the color excess E(B-V) depending on the line of sight through the cluster. }
\label{fig:ML_jerk}
\end{figure}

Figure \ref{fig:ML_jerk} shows the measured cluster mass and V-band luminosity using the cluster parameters obtained from the measurements of acceleration and jerk. We find that the integrated mass along our line of sight within a cylindrical radius of 1 parsec to be M($R_\perp<$1pc)$\sim$3.0$\times$10$^5$ M$_\odot$. We also find a similar central mass to light ratio in the core of M/L$_v\simeq$2.0$_{-0.7}^{+0.8}$ when compared to the acceleration only measurements. Again, we note that the errors in the exact reddening towards Terzan 5 produce significant systematic offsets in these results.

The predicted root mean square velocity of a star in the core of a King model cluster is given by:
\begin{equation}
v_c = \left(\frac{4\pi}{3}G\rho_c\right)^{1/2}r_c\,.
\label{eqn:los_vel}
\end{equation}

\begin{figure}[t]
  \centering
  \includegraphics[width=\columnwidth]{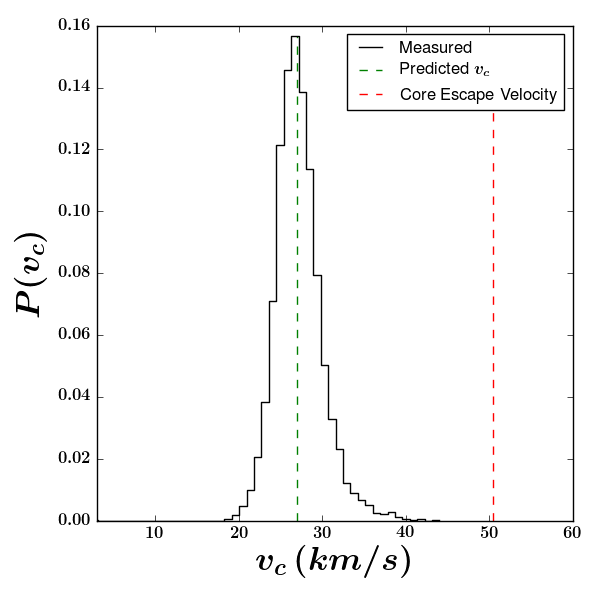}
  \caption
  { Distribution of measured velocities for all pulsars and across all chains. We plot the expected root mean square core velocity for the pulsars given by Equation \ref{eqn:los_vel} in green. We find that the pulsars follow an approximately Gaussian distribution about these predicted values. The escape velocity needed to escape from the core of the cluster is plotted in red \citep{2002ApJ...568L..23G}. }
\label{fig:velocity_dist}
\end{figure}

Figure \ref{fig:velocity_dist} shows the distribution of all line of sight pulsar velocities over all chains for a model with flat priors in our velocity fitting. We find our models predict a roughly Gaussian distribution of line of sight velocities about the predicted value, confirming our jerk measurements are drawn from a cluster potential that obeys the King model.

We find a distribution of velocities for our pulsars centered about $v_c$=$\pm$26.3$_{-2.7}^{+3.6}$ km/s, as shown in Figure \ref{fig:velocity_dist}. The one-dimensional velocity dispersion was measured to be $\sigma$=$\pm$15.2$_{-1.5}^{+2.1}$ km/s.

\citet{2013ApJ...779L...5O} gives an estimate of the velocity dispersion for the most metal poor population in Terzan 5 of $\sigma\sim15$ km/s. We verify our results by finding the velocity predicted by the \citet{2013ApJ...779L...5O} results if their population of stars are in equipartition with the pulsars and have $q$=$\qaccel$. We find that the results of \citet{2013ApJ...779L...5O} predict a pulsar velocity dispersion of $\sim$12 km/s, which is similar to our results. A more detailed study of the velocity dispersion of both the pulsars and the main sequence turn-off stars is required to further confirm this result.

From \citet{2002ApJ...568L..23G} we find that the predicted escape velocity for Terzan 5 from the core of the cluster is 50.5 km/s. Figure \ref{fig:velocity_dist} shows the escape velocity in red, a few standard deviations from the median of the observed distribution.

\subsection{Center of gravity fits}
\label{subsec:cog_results}

\begin{figure}[t]
  \centering
  \includegraphics[width=\columnwidth]{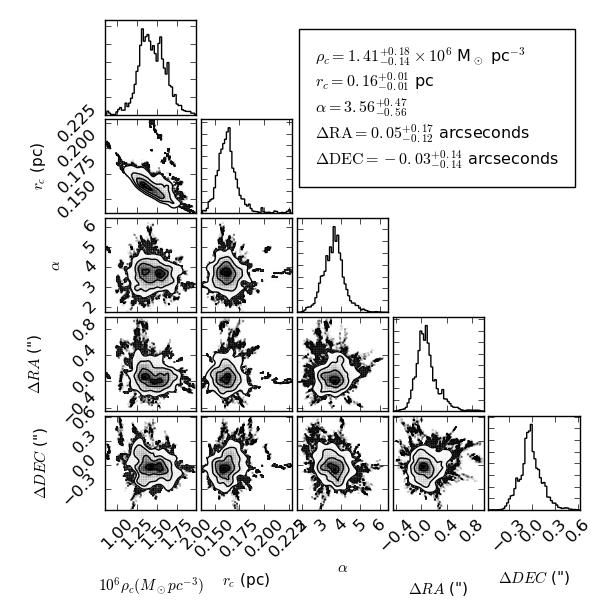}
  \caption
  { Corner plot showing the marginalized probabilities of finding a given cluster parameter for the cluster Terzan 5 with an allowed offset from the optical CoG as found by \citet{2012ApJ...755L..32M} including the covariances between different parameters. These results use only the pulsar accelerations. }
\label{fig:cog_corner}
\end{figure}

The results of our fits for the cluster parameters $\rho_c$, $r_c$, and $\alpha$ including an offset in the CoG of the cluster using both pulsar accelerations and jerks are shown in Figure \ref{fig:cog_corner}. Our analysis finds results consistent with the optical CoG published by \citet{2012ApJ...755L..32M}, with a posterior distribution centered on $\Delta$RA=0.05$_{-0.12}^{+0.17}$ arcseconds and $\Delta$Dec=-0.03$_{-0.14}^{+0.14}$ arcseconds is much more constraining than our prior.

\subsection{Central black hole results}
\label{subsec:accel_bh_results}

\begin{figure}[t]
  \centering
  \includegraphics[width=\columnwidth]{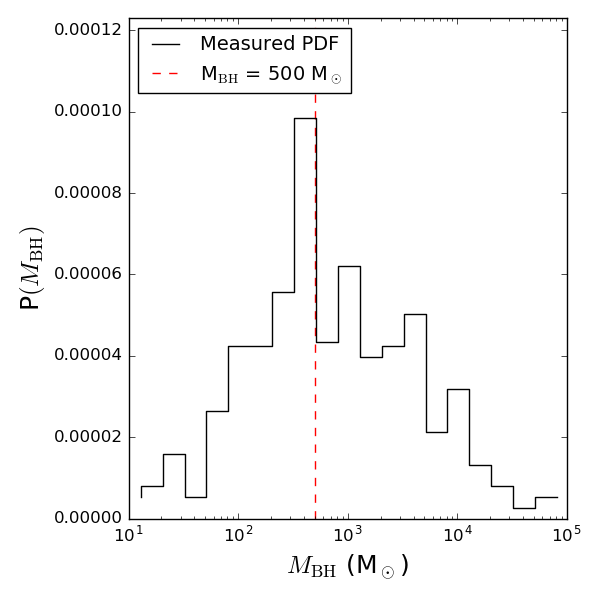}
  \caption
  { Observed posterior distribution for the central black hole mass in Terzan 5. We find a marginal detection of a black hole mass of $\mbhcand$ using the measured accelerations and jerks. We also find that black hole masses greater than $\mbhupper$ are ruled out by our simulations. }
\label{fig:bh_mass_pdf}
\end{figure}

The results of our black hole fits are shown in Figure \ref{fig:bh_mass_pdf}. We find that 99\% of all the chains in our simulation favor a black hole mass below $\mbhupper$. At the intermediate values of a few hundred solar masses however, we find a marginal detection for a black hole of mass $\mbhcand$. We stress that additional analysis is required to confirm the existence of a black hole of this mass. More constraining fits could be accomplished by either finding new pulsars near to the core of the cluster or by implementing a more robust black hole model that allows the black hole to be offset from the CoG and accounting for the jerks produced by the black hole.

An upper limit on black hole mass of $\mbhupper$ is still within the bounds of current theory for hierarchical black hole formation in a natal globular cluster. Simulations by \citet{2004ApJ...613.1143B} have found that at the end of growth of the black hole for these systems, the expected black hole masses according to their simulations should be between a few hundred and a thousand M$_\odot$. We also compare our results with those of \citet{2010ApJ...710.1063V}, which find an upper limit on a central black hole mass of 10$^4$ M$_\odot$ for $\omega$ Centauri. 

\section{Discussion}
\label{sec:discussion}

\subsection{Terzan 5 Taxonomy}
\label{subsec:discussion_taxonomy}

The primary goal of this work has been to derive a reddening independent way to study the cluster parameters for Terzan 5 using an MCMC sampler to assign a proper location and spin-down rate to each pulsar in the system. This was accomplished by assuming the cluster obeys the King model \citep{1962AJ.....67..471K} and by varying the exact spectral index of the pulsar density distribution to be relative to the dominant mass class in the core of the cluster. Studying Terzan 5 in this manner is allows us a direct probe of the gravitating mass of the cluster, and does not make any assumptions about the mass to light ratio of the system.

As our models are only valid within the core, we cannot derive a total mass out to the tidal radius of Terzan 5. Optical studies have found the system to be highly concentrated however \citep{2010ApJ...717..653L}, therefore we expect that our integrated mass out to a few core radii represents a majority of the mass in the system and is indicative of a lower total mass than found by \citet{2010ApJ...717..653L}. From the results of \citet{1987IAUS..125..187V}, we find the collision rate ($\Gamma\propto\rho_c^{3/2}r_c^{2}$) of Terzan 5 to be slightly smaller than the results of \citet{2010ApJ...717..653L} ($\sim70\%$). Our results are still a factor of a few larger than those measured for other massive globular clusters however, which maintains Terzan 5 distinction for having the largest measured collision rate of any globular cluster studied to date.

We find that the total mass and the mass to light ratios place strong constraints on the possible taxonomy of Terzan 5. \citet{2009Natur.462..483F} suggested that Terzan 5 might be the nuclear remnant of a dwarf galaxy. Without detailed knowledge of the original galactocentric distance, we use the results of \citet{1999MNRAS.302..771J} to estimate that the current mass of a dwarf galaxy over the past 10 Gyrs to be $<$1\% of its original value. If Terzan 5 was a dwarf galaxy at its inception, then its current mass is on the lower end of the commonly accepted mass range for dwarf galaxies of 10$^7$ $-$ 10$^9$ M$_\odot$ \citep{2015ApJ...805....2S} after tidal stripping. In addition to this, the mass to light ratio of Milky Way ultra-faint satellites are believed to be a few hundred \citep{2007ApJ...670..313S}. Even accounting for tidal stripping as the progenitor galaxy enters the Milky-Way potential, the required change in the mass to light ratio is likely far too large for Terzan 5 to have been a dwarf galaxy at its formation.

\citet{2009Natur.462..483F} also predicted that Terzan 5 may be a possible fragment of the Milky Way's bulge. For Milky Way-like spiral galaxies, \citet{2008PASJ...60..493Y} measured average mass to light ratios much smaller than those of dwarf galaxies. While we have a large systematic uncertainty in our measured mass to light ratio for Terzan 5, our results agree much better with this formation mechanism as the maximum mass to light ratio will always be much less than that of a disrupted dwarf galaxy.

\subsection{Further improvements}
\label{subsec:discussion_improvements}

We can make future improvement to our method by moving away from a simple King model \citep{1962AJ.....67..471K} to a model that accounts for the underlying IMF of the system and better accounts for the entire range of masses in the cluster. At this time our work only approximates a multi-mass system by referencing the pulsars to the dominant mass class in the core. Using a distribution function which accounts for the mass of each star in the globular cluster as well as its position and velocity, each step of our model would be able to more accurately determine the effect neighboring stars have on the pulsars.

It should also be noted that with first light from the next generation of radio telescopes such as MeerKAT \citep{2009arXiv0910.2935B}, the number of timed pulsars in globular clusters, particular those in the southern hemisphere, will increase and allow us to probe the potential of more globular clusters using pulsar timing.

\subsection{Method Comparison}

By assigning line of sight positions and spin-down rates for each pulsar at each possible set of cluster parameters, our work gains an important distinction from previous globular cluster studies using pulsars. Models produced using only the on-sky positions and Equation \ref{eqn:phin_amax} suffer from a strong covariance in the core density and core radius. Using the joint probability of the line of sight position for each pulsar and its expected acceleration, we can produce models that are far less covariant than previous works, as evidenced by comparing Figures \ref{fig:loglike} and \ref{fig:accel_only_corner}.

\subsection{Mass Ratios}

As a result of determining the density profile for Terzan 5, we also infer the underlying system mass for the $\Npsr$ pulsars in our study. We have found a mass ratio of pulsars to the dominant mass class of q=$\qaccel$. As we expect pulsars to follow the same mass distribution as LMXBs (inset plot of Figure \ref{fig:cum_dist}), we compare this to the results of \citet{2006ApJ...651.1098H} ($\qheinke$), finding good agreement between the two results. For a turn-off mass of 0.9 M$_\odot$ \citep{2001ApJ...556..322B} this predicts a system mass for the neutron star population in this study of $\sim$1.3 M$_\odot$.

Comparing this to the work done by \citet{2013ApJ...778...66K}, \citet{2016arXiv160302698O}, and \citet{2016arXiv160501665A}, we should expect the neutron star mass distribution to be broadly centered on 1.55$_{-0.20}^{+0.26}$ M$_\odot$. This is larger than our modeled system masses in our MSP population, which must also include the unknown mass of the white dwarf companion present in most of our pulsar systems.

Two possible explanations for this discrepancy can be related back to our assumptions of the dominant mass class in the core of the cluster. The first is that should the core of Terzan 5 have a large population of heavy stellar remnants and binaries, the resulting system mass of the pulsars could be much closer to the expected neutron star mass range. These heavy stellar systems could be comprised of such systems as heavy white dwarfs or hard binaries. The presence of hard binaries, which is a binary system with a binding energy greater than the average for other binaries in the cluster, can halt gravothermal collapse of the system \citep{1975MNRAS.173..729H} and is an ideal means of explaining why Terzan 5 has avoided collapse. While the IMF of the system will not allow for large numbers of heavy stellar remnants to be formed, mass segregation would rapidly produce an overpopulation of these systems in the core compared to the entirety of the cluster. If these stars are the dominant mass class of the cluster, they too would provide a much more reasonable system mass for the pulsars according to our measured mass ratio.

The second explanation is that the pulsars in Terzan 5 were formed via an electron capture supernovae. This argument also agrees with the hypothesis that the core may have an abundance of heavy white dwarfs, as these are believed to have been the possible progenitors for the Terzan 5 pulsars \citep{2008MNRAS.386..553I}. Should this be the case, evidence exists that the median mass distribution of neutron stars formed via electron capture is $\sim1.25$ M$_\odot$. If the pulsar binaries in our sample are orbitted by a white-dwarf of a few tenths of a solar mass, this agrees well with our measurements.

\subsection{Black Hole Fits}

To date, Terzan 5 is one of the best candidates for testing the existence of a central massive black hole given its large number of confirmed pulsars deep within its core. In order to expand this work to other globular clusters, and to improve our limits on the black hole mass for Terzan 5, we would either need to increase the number of available pulsars in each globular cluster or create a more robust model for the physics governing the black hole.

We argue that increasing the number of known pulsars within a globular cluster can improve our constraints on the presence of a black hole for two separate reasons. The first is that the more pulsars we are able to sample the potential with, the more constraints we can put on the accelerations of each pulsar. We are therefore better able to discern if a pulsar is truly better fit by being within, or nearly within, the radius of influence of the black hole. We also argue that the likelihood of finding a pulsar that is very centrally concentrated is increased, which gives us the best constraints on the possible upper mass limit of a black hole. Finding these new pulsars is highly likely as the next generation of radio telescopes turn on and the sensitivity limit for finding weak pulsars is improved \citep{2015aska.confE..47H}. It is also expected that search methods for finding weak pulsars will improve over time, as evidenced by the discovery of two new (but as of yet not well timed enough to include in this work) pulsars in 47 Tucanae \citep{2016MNRAS.459L..26P} and the three new pulsars discovered in (one of which does not have a timing solution) in Terzan 5 (Cadelano et al. 2017). 

The latter method of improving our model for the black hole physics may be achieved by allowing the black hole to be offset from the center of the cluster and incorporating its contribution to the jerk of each pulsar. As the black holes mass is much less than the total mass of the cluster, we do not expect it to reside at the exact CoG of the system. Incorporating a positional offset in the black hole's position from the CoG may allow for individual pulsars to be closer or further from the black hole, thereby experiencing accelerations and jerks that cannot be explained by the King model.

Without these changes, we can only take our results as being suggestive in the singular case of Terzan 5 that there is an upper limit to the mass of an intermediate mass black hole in the system.

\section{Summary}

Using the MCMC sampler {\tt emcee} \citep{2013PASP..125..306F}, we have used 36 MSPs within the globular cluster Terzan 5 to better constrain the cluster parameters, including the core density $\rho_c$=$\dencjerk\times10^6$ M$_\odot$ pc$^{-3}$, the core radius $r_c$=$\rcjerk$ pc, and the spectral index of the neutron star density $\alpha$=$\alphajerk$ by using a reddening independent method that probes the true mass density profile and does not rely on an assumed mass to light ratio. These values agree well with previously obtained quantities using HST and ESO-MAD, while using a method that allows us to avoid the issue of reddening or stellar counting towards the cluster center. This is a direct improvement on the previous methods of studying cluster dynamics via pulsar timing as we can better determine the line of sight position of each pulsar in the cluster. 

As our density profile for the system is completely derived from the gravitating mass of the system, we are able to provide a more sensitive probe of the true mass of the system. Our methods find an integrated mass of M$_T\simeq$3.5$\times$10$^5$ M$_\odot$ along our line of sight and within an impact parameter of 1 parsec of the core. Without a better luminosity profile however, we cannot provide better estimates for the true mass to light ratio of the system. The best estimate for the mass to light ratio for the system is M/L$_v\simeq$2.0$_{-0.7}^{+0.8}$ when using all of the pulsar timing data available, though this quantity is dominated by systematics due to reddening effects.

We find our method for studying the Terzan 5 cluster dynamics provides better evidence that Terzan 5 is likely a pristine fragment of the Galactic bulge \citep{2009Natur.462..483F} and not the disrupted nucleus of a dwarf galaxy that entered the tidal field of the Milky Way. We also find evidence that the core of the system may be comprised primarily of heavy stellar remnants and binaries, which could have provided the needed evolutionary channel to produce the large population of MSPs observed via accretion induced collapse. Finally we find that Terzan 5 has an upper limit on the central black hole mass of $\mbhupper$. Future work will better constrain the black hole mass of this system using more sophisticated acceleration models for the pulsar timing in order to help better constrain the formation history of the Terzan 5 core.

\section{Acknowledgements}

Observations were taken with the Green Bank Telescope, which are operated by the National Radio Astronomy Observatory, a facility of the National Science Foundation operated under cooperative agreement by Associated Universities, Inc.

B.P. acknowledges funding from the University of Virginia Astronomy department, the Grote Reber fellowship through the National Radio Astronomy Observatory, and the Graduate Research STEM Fellowship Program through the Virginia Space Grant Consortium under the National Aeronautics and Space Administration (NASA). 

S.R. is also an staff Astronomer at the National Radio Astronomy Observatory.

J.W.T.H. acknowledges funding from an NWO Vidi fellowship and from the European Research Council under the European Union's Seventh Framework Programme (FP/2007-2013) / ERC Starting Grant agreement nr. 337062 ("DRAGNET").

I.S. acknowledges pulsar research at UBC is funded by an NSERC Discovery Grant and by the Canadian Institute for Advanced Research.

P.A. is supported by NASA Origins Grant NNX14AE16G and NASA ATP grant NNH12ZDA.

\bibliographystyle{apj}

\clearpage
\begin{deluxetable*}{rrrrrrrrr}
 \tablecaption{Timing Parameters for Terzan 5 Pulsars}
 \tablehead{
   \colhead{PSR} & \colhead{RA} & \colhead{DEC} & \colhead{$P$} & \colhead{$\dot{P}$} & \colhead{$\ddot{P}$} & \colhead{$P_b$} & \colhead{$\dot{P}_b$} & \colhead{Duration} \\
   \colhead{} & \colhead{(17:48:)} & \colhead{($-$24:46:)} & \colhead{(ms)} & \colhead{($10^{-19}$ s s$^{-1}$)} & \colhead{($10^{-30}$ s$^{-1}$)} & \colhead{(days)} & \colhead{($10^{-12}$ s s$^{-1}$)} & \colhead{(yrs)} } \\
\startdata
A$^1$ & 02.251 & 37.37 & 12.00 & -0.29 & $-$ & 0.08 & $-$ & 1.65 \\ 
C & 04.534 & 34.72 & 8.40 & -6.00 & 0.52 & $-$ & $-$ & 17.67 \\ 
D & 05.922 & 05.67 & 4.70 & 1.30 & -0.16 & $-$ & $-$ & 17.67 \\ 
E & 03.409 & 35.48 & 2.20 & -0.18 & 2.00 & 60.00 & $-$ & 14.84 \\ 
F & 05.118 & 38.05 & 5.50 & 0.04 & -5.10 & $-$ & $-$ & 14.84 \\ 
G & 05.650 & 46.59 & 22.00 & 3.90 & -7.70 & $-$ & $-$ & 14.84 \\ 
H & 05.634 & 53.06 & 4.90 & -0.83 & -0.74 & $-$ & $-$ & 14.84 \\ 
I & 04.874 & 46.37 & 9.60 & -0.71 & -26.00 & 1.30 & -3.00 & 17.67 \\ 
J & 04.010 & 1:40.33 & 80.00 & 25.00 & -8.20 & 1.10 & -1.30 & 17.67 \\ 
K & 03.909 & 47.73 & 3.00 & -0.94 & -0.26 & $-$ & $-$ & 14.84 \\ 
L & 04.738 & 35.81 & 2.20 & -0.17 & 1.10 & $-$ & $-$ & 14.84 \\ 
M & 04.618 & 40.75 & 3.60 & 4.90 & 3.30 & 0.44 & 5.20 & 14.84 \\ 
N & 04.919 & 53.78 & 8.70 & 5.50 & 2.00 & 0.39 & 2.00 & 17.67 \\ 
O & 04.682 & 51.40 & 1.70 & -0.69 & 2.10 & 0.26 & -1.50 & 11.26 \\ 
P$^1$ & 05.038 & 41.36 & 1.70 & 2.60 & $-$ & 0.36 & $-$ & 3.73 \\ 
Q & 04.336 & 1:05.03 & 2.80 & -0.36 & 0.19 & 30.00 & $-$ & 11.26 \\ 
R & 04.688 & 50.25 & 5.00 & 4.70 & -18.00 & $-$ & $-$ & 14.84 \\ 
S & 04.293 & 31.71 & 6.10 & 0.64 & 0.69 & $-$ & $-$ & 14.84 \\ 
T & 02.991 & 52.81 & 7.10 & 3.10 & 0.71 & $-$ & $-$ & 14.84 \\ 
U & 04.244 & 47.86 & 3.30 & 3.00 & 0.01 & 3.60 & 29.00 & 11.26 \\ 
V & 05.106 & 34.46 & 2.10 & -0.95 & 0.17 & 0.50 & -2.20 & 11.26 \\ 
W & 04.840 & 42.38 & 4.20 & 1.20 & -18.00 & 4.90 & 9.30 & 14.84 \\ 
X & 05.594 & 1:12.06 & 3.00 & 0.59 & 0.04 & 5.00 & 7.40 & 14.84 \\ 
Y & 05.097 & 44.57 & 2.00 & 1.50 & -120.00 & 1.20 & 6.40 & 11.26 \\ 
Z & 04.946 & 46.11 & 2.50 & -0.86 & 17.00 & 3.50 & -11.00 & 14.84 \\ 
aa & 05.812 & 42.24 & 5.80 & -4.40 & 0.91 & $-$ & $-$ & 14.84 \\ 
ab & 04.759 & 42.65 & 5.10 & 4.20 & -3.10 & $-$ & $-$ & 14.84 \\ 
ac & 06.044 & 32.53 & 5.10 & 2.30 & 1.20 & $-$ & $-$ & 11.26 \\ 
ad$^1$ & 03.847 & 41.86 & 1.40 & -0.34 & $-$ & 1.10 & $-$ & 4.16 \\ 
ae & 04.964 & 45.72 & 3.70 & -5.70 & -2.80 & 0.17 & -2.80 & 11.26 \\ 
af & 04.212 & 44.87 & 3.30 & -2.30 & -0.21 & $-$ & $-$ & 11.26 \\ 
ag & 04.811 & 34.59 & 4.40 & 0.12 & -1.20 & $-$ & $-$ & 11.26 \\ 
ah & 04.321 & 42.03 & 5.00 & 5.70 & 0.60 & $-$ & $-$ & 11.26 \\ 
ai & 04.115 & 50.34 & 21.00 & 14.00 & 6.90 & 0.85 & 3.80 & 11.26 \\ 
aj & 05.012 & 34.69 & 3.00 & 1.40 & 1.70 & $-$ & $-$ & 11.96 \\ 
ak & 03.686 & 37.93 & 1.90 & 0.88 & 0.49 & $-$ & $-$ & 11.45
\enddata
\label{table:Params}
\tablecomments{Timing parameters for the $\Npsr$ pulsars in Terzan 5 used for this study. No pulsar with the name ``B'' exists within Terzan 5, as the pulsar originally given this designation was shown to be unassociated with Terzan 5 \citep{2000MNRAS.316..491L}. Right ascension and declination for each pulsar is recorded in seconds of time and degree respectively. The duration for these timing models span between 1.5 to 15 years, with most models being accurate for approximately eleven years of data. \\ $^{1}$ - Redback pulsars. These systems experience various forms of timing noise which limit the quality of higher order spin period derivatives and greatly complicates the creation of a long-term phase connected timing solution. We therefore do not show the orbital period derivatives for these systems as they track different physics that is unrelated to the cluster dynamics. }
\end{deluxetable*}

\begin{deluxetable*}{lrrrrrrrrrrr}
 \tablecaption{Black Widow \& Pulsar-White Dwarf Binary Orbital Properties}
 \tablehead{
   \colhead{PSR} & \colhead{} & \colhead{} & \colhead{} & \colhead{$c\left|\dot{P}_b/P_b\right|$} & \colhead{} & \colhead{} & \colhead{} & \colhead{B} &  \colhead{} & \colhead{} & \colhead{Association} \\[2.5mm] 
   \colhead{} & \colhead{} & \colhead{} & \colhead{} & \colhead{(m s$^{-2}$)} & \colhead{} & \colhead{} & \colhead{} & \colhead{(10$^8$ G)} & \colhead{} & \colhead{} & \colhead{} } \\
\startdata
\multicolumn{12}{c}{} \\
\multicolumn{12}{c}{\emph{Black Widows}} \\
\hline \\
J2051$-$0827$^\textbf{*}$ & & & & 5.4$\times$10$^{-7}$ & & & & - & & & Field$^1$ \\
J1959$+$2048$^\textbf{*}$ & & & & 1.3$\times$10$^{-7}$ & & & & - & & & Field$^2$ \\
J1731$-$1847$^\textbf{*}$ & & & & 1.2$\times$10$^{-7}$ & & & & - & & & Field$^3$ \\
J0021$-$7204I & & & & 6.1$\times$10$^{-9}$ & & & & 5.7 & & & 47 Tuc$^4$ \\
J0021$-$7204J & & & & 1.5$\times$10$^{-8}$ & & & & 8.7 & & & 47 Tuc$^4$ \\
J0021$-$7204O$^\textbf{*}$ & & & & 2.3$\times$10$^{-7}$ & & & & $^\textbf{**}$ & & & 47 Tuc$^4$ \\
J1748$-$2446ae & & & & 5.7$\times$10$^{-8}$ & & & & 12.0 & & & Terzan 5$^5$ \\
J1748$-$2446O & & & & 2.0$\times$10$^{-8}$ & & & & 5.2 & & & Terzan 5 \\
\multicolumn{12}{c}{} \\
\multicolumn{12}{c}{\emph{Pulsar-White Dwarfs}} \\
\hline \\
J1748$-$2446M & & & & 4.0$\times$10$^{-10}$ & & & & 2.5 & & & Terzan 5 \\
J1748$-$2446N & & & & 1.5$\times$10$^{-9}$ & & & & 12.1 & & & Terzan 5 \\
J1748$-$2446V & & & & 1.5$\times$10$^{-9}$ & & & & 2.8 & & & Terzan 5 \\
J1748$-$2446W & & & & 2.1$\times$10$^{-9}$ & & & & 6.9 & & & Terzan 5 \\
J1748$-$2446X & & & & 7.0$\times$10$^{-10}$ & & & & 2.8 & & & Terzan 5 \\
J1748$-$2446Y & & & & 3.6$\times$10$^{-9}$ & & & & 4.4 & & & Terzan 5 \\
J1748$-$2446Z & & & & 3.8$\times$10$^{-10}$ & & & & 1.7 & & & Terzan 5$^5$ \\
\enddata
\label{table:bws}
\tablecomments{ Black widows and pulsar-white dwarfs with measured values of $\dot{P}_b$ in the field of the galaxy, 47 Tucanae, and Terzan 5. The three field black widows have been observed to have orbital phase wander. Of the five globular cluster black widows, one pulsar (J0021$-$7204O) has confirmed orbital phase wander. For the pulsars in globular clusters we have also shown the derived magnetic field strength according to the derived value of $\dot{P}$ given by Equation \ref{eq:ae_o_pdot}, of which only J0021$-$7204O appear to be anomalous due to its confirmed orbital phase wander$^4$. We find similar magnetic field strengths between the black widows and pulsar-white dwarf systems. \\
$^\textbf{*}$ - Has been confirmed to have orbital phase wander. \\
$^\textbf{**}$ - Due to phase wander, attempts to measure a magnetic field strength yield an non-physical negative value. \\
$^1$ - \citet{2011MNRAS.414.3134L} \\ 
$^2$ - \citet{1990ApJ...351..642F} \\ 
$^3$ - \citet{2011MNRAS.416.2455B} \\ 
$^4$ - \citet{2003MNRAS.340.1359F} \\ 
$^5$ - \citet{Ter5timingprep} }
\end{deluxetable*}

\begin{deluxetable*}{lrrr}
 \tablecaption{Cluster Parameters for Terzan 5}
 \label{table:accel_params}
 \tablehead{
   \colhead{Fitting Method} & \colhead{$\rho_c$} & \colhead{$r_c$} & \colhead{$\alpha$} \\
   \colhead{} & \colhead{(10$^6$ M$_\odot$pc$^{-3}$)} & \colhead{(pc)} & \colhead{} }\\
\startdata
Previously Cited & (1-4)$^1$ & 0.16$\pm$0.02$^2$ & $-$3.29$\pm$0.33$^3$ \\[1.5mm]
Projected & $\dencinit$ & $\rcinit$ & $\alphainit$ \\[1.5mm]
Accelerations Only & $\dencaccel$ & $\rcaccel$ & $\alphaaccel$ \\[1.5mm]
Accelerations \& Jerks & $\dencjerk$ & $\rcjerk$ & $\alphajerk$ \\[1.5mm]
Accelerations \& Jerks \& Black Hole Fitting & $\dencBH$ & $\rcBH$ & $\alphaBH$ \\[1.5mm]
Accelerations \& Jerks \& Center of Gravity Offset & $\dencCOG$ & $\rcCOG$ & $\alphaCOG$ \\[1.5mm]
\enddata
\label{table:cluster_params}
\tablecomments{ Cluster parameters for Terzan 5 using different fitting methods. \\
$^1$ - \citet{2010ApJ...717..653L} \\ 
$^2$ - \citet{2013ApJ...774..151M}: Main sequence turn off star core radius scaled down to the neutron star core radius according to Equation \ref{eq:mass_relations_king} using the measured spectral index of \citet{2006ApJ...651.1098H} \\
$^3$ - \citet{2006ApJ...651.1098H} }
\end{deluxetable*}

\begin{deluxetable*}{lrrrrrrrrr}
 \tablecaption{Line of sight positions using projected cluster parameters \\[1.5mm] $\rho_c=\dencinit\times10^6$ M$_\odot$ pc$^{-3}$ \\ $r_c=\rcinit$ pc \\ $\alpha=\alphainit$ \\ d = 5.9 kpc }
 \label{table:accel_params}
 \tablehead{
   \colhead{PSR} & \colhead{$R_\perp$} & \colhead{$a_{\rm z,meas}$} & \colhead{$l_1$} & \colhead{$l_2$} & \colhead{$r_1$} & \colhead{$r_2$} & \colhead{$P(l_1)$} & \colhead{$P(l_2)$} & \colhead{$\frac{P(l_1)}{P(l_2)}$} \\
   \colhead{} & \colhead{(pc)} & \colhead{($10^{-9}$ m s$^{-2}$)} & \colhead{(pc)} & \colhead{(pc)} & \colhead{(pc)} & \colhead{(pc)} & \colhead{} & \colhead{} & \colhead{} } \\
\startdata
I & 0.052 & $-$7.779 & 0.009 & 1.390 & 0.052 & 1.391 & 3.82e$-$18 & 4.61e$-$18 & 0.83 \\ 
ae & 0.055 & $-$56.908 & 0.079 & 0.311 & 0.096 & 0.315 & 2.86e$-$17 & 2.90e$-$17 & 0.99 \\ 
Z & 0.057 & $-$10.795 & 0.013 & 1.138 & 0.059 & 1.140 & 5.41e$-$18 & 6.14e$-$18 & 0.88 \\ 
W & 0.063 & 6.632 & $-$0.008 & $-$1.527 & 0.064 & 1.528 & 3.39e$-$18 & 4.02e$-$18 & 0.84 \\ 
ab & 0.066 & 24.510 & $-$0.030 & $-$0.660 & 0.073 & 0.663 & 1.27e$-$17 & 1.30e$-$17 & 0.98 \\ 
Y & 0.096 & 19.053 & $-$0.027 & $-$0.782 & 0.100 & 0.788 & 1.15e$-$17 & 1.04e$-$17 & 1.11 \\ 
P & 0.118 & 44.924 & $-$0.083 & $-$0.374 & 0.145 & 0.392 & 2.96e$-$17 & 2.47e$-$17 & 1.19 \\ 
M & 0.142 & 41.097 & $-$0.088 & $-$0.396 & 0.168 & 0.421 & 3.07e$-$17 & 2.34e$-$17 & 1.31 \\ 
R & 0.173 & 28.264 & $-$0.069 & $-$0.553 & 0.187 & 0.579 & 2.60e$-$17 & 1.62e$-$17 & 1.60 \\ 
O & 0.205 & $-$20.030 & 0.058 & 0.717 & 0.213 & 0.745 & 2.28e$-$17 & 1.16e$-$17 & 1.96 \\ 
F & 0.214 & 0.194 & $-$0.001 & $-$10.484 & 0.214 & 10.486 & 2.38e$-$19 & 2.28e$-$19 & 1.04 \\ 
ah & 0.218 & 34.654 & $-$0.131 & $-$0.407 & 0.255 & 0.462 & 3.63e$-$17 & 2.28e$-$17 & 1.59 \\ 
af & 0.248 & $-$20.670 & 0.081 & 0.673 & 0.261 & 0.717 & 2.91e$-$17 & 1.26e$-$17 & 2.31 \\ 
U & 0.254 & 28.542 & $-$0.130 & $-$0.479 & 0.285 & 0.542 & 3.63e$-$17 & 1.91e$-$17 & 1.90 \\ 
L & 0.255 & $-$2.256 & 0.009 & 2.807 & 0.255 & 2.818 & 3.70e$-$18 & 1.64e$-$18 & 2.26 \\ 
N & 0.264 & 17.625 & $-$0.075 & $-$0.758 & 0.275 & 0.803 & 2.77e$-$17 & 1.08e$-$17 & 2.57 \\ 
ag & 0.287 & 0.839 & $-$0.004 & $-$4.826 & 0.287 & 4.834 & 1.67e$-$18 & 7.30e$-$19 & 2.29 \\ 
aj & 0.290 & 14.275 & $-$0.071 & $-$0.879 & 0.299 & 0.926 & 2.64e$-$17 & 8.83e$-$18 & 2.99 \\ 
V & 0.306 & $-$15.157 & 0.084 & 0.828 & 0.318 & 0.883 & 2.97e$-$17 & 9.58e$-$18 & 3.10 \\ 
C & 0.308 & $-$21.474 & 0.131 & 0.597 & 0.335 & 0.672 & 3.63e$-$17 & 1.47e$-$17 & 2.47 \\ 
G & 0.317 & 5.449 & $-$0.030 & $-$1.668 & 0.318 & 1.698 & 1.27e$-$17 & 3.53e$-$18 & 3.60 \\ 
ai & 0.330 & 15.550 & $-$0.100 & $-$0.794 & 0.345 & 0.860 & 3.29e$-$17 & 1.01e$-$17 & 3.24 \\ 
K & 0.377 & $-$9.501 & 0.076 & 1.133 & 0.385 & 1.194 & 2.78e$-$17 & 6.18e$-$18 & 4.50 \\ 
aa & 0.381 & $-$22.737 & 0.284 & 0.408 & 0.475 & 0.558 & 3.09e$-$17 & 2.27e$-$17 & 1.36 \\ 
H & 0.390 & $-$5.084 & 0.042 & 1.716 & 0.392 & 1.760 & 1.73e$-$17 & 3.39e$-$18 & 5.11 \\ 
ad & 0.398 & $-$7.273 & 0.064 & 1.356 & 0.403 & 1.413 & 2.46e$-$17 & 4.78e$-$18 & 5.14 \\ 
S & 0.428 & 3.150 & $-$0.032 & $-$2.282 & 0.429 & 2.322 & 1.33e$-$17 & 2.23e$-$18 & 5.95 \\ 
ak & 0.492 & 14.025 & $-$0.241 & $-$0.690 & 0.547 & 0.847 & 3.41e$-$17 & 1.22e$-$17 & 2.79 \\ 
ac & 0.579 & 13.428 & $-$0.479 & $-$0.479 & 0.752 & 0.752 & 1.91e$-$17 & 1.91e$-$17 & 1.00 \\ 
Q & 0.618 & $-$3.879 & 0.091 & 1.931 & 0.625 & 2.028 & 3.11e$-$17 & 2.85e$-$18 & 10.90 \\ 
E & 0.619 & $-$2.450 & 0.056 & 2.583 & 0.621 & 2.657 & 2.22e$-$17 & 1.85e$-$18 & 12.00 \\ 
T & 0.761 & 13.067 & $-$0.608 & $-$0.608 & 0.974 & 0.974 & 1.44e$-$17 & 1.44e$-$17 & 1.00 \\ 
X & 0.837 & 5.156 & $-$0.282 & $-$1.384 & 0.884 & 1.617 & 3.11e$-$17 & 4.64e$-$18 & 6.70 \\ 
A & 1.033 & $-$0.745 & 0.061 & 4.995 & 1.035 & 5.101 & 2.37e$-$17 & 6.93e$-$19 & 34.20 \\ 
D & 1.189 & 8.074 & $-$0.914 & $-$0.914 & 1.500 & 1.500 & 8.37e$-$18 & 8.37e$-$18 & 1.00 \\ 
J & 1.627 & $-$4.087 & 1.228 & 1.228 & 2.039 & 2.039 & 5.51e$-$18 & 5.51e$-$18 & 1.00 \\  
\enddata
\label{table:initial_pos}
\tablecomments{ Solutions for the line of sight positions of each pulsar using the cluster parameters derived using traditional parameter fitting methods. Pulsars are ordered by their projected distance from the centers of the cluster. }
\end{deluxetable*}

\begin{deluxetable*}{lrrrrrr}
 \tablecaption{Line of sight positions}
 \label{table:accel_results}
 \tablehead{\colhead{PSR} & \colhead{Accelerations} & \colhead{} & \colhead{} & \colhead{Acceleration \& Jerk} & \colhead{} & \colhead{}\\
 \colhead{} & \colhead{$l_1$ (pc)} & \colhead{$l_2$ (pc)} & \colhead{$\frac{P\left(l_1\right)}{P\left(l_2\right)}$} & \colhead{$l_1$ (pc)} & \colhead{$l_2$ (pc)} & \colhead{$\frac{P\left(l_1\right)}{P\left(l_2\right)}$} } \\
\startdata
I & $\zprimaryaccelI$ & $\zsecondaryaccelI$ & $\zprobaccelI$ & $\zprimaryjerkI$ & $\zsecondaryjerkI$ & $\zprobjerkI$ \\[1.5mm]
ae & $\zprimaryaccelae$ & $\zsecondaryaccelae$ & $\zprobaccelae$ & $\zprimaryjerkae$ & $\zsecondaryjerkae$ & $\zprobjerkae$ \\[1.5mm]
Z & $\zprimaryaccelZ$ & $\zsecondaryaccelZ$ & $\zprobaccelZ$ & $\zprimaryjerkZ$ & $\zsecondaryjerkZ$ & $\zprobjerkZ$ \\[1.5mm]
W & $\zprimaryaccelW$ & $\zsecondaryaccelW$ & $\zprobaccelW$ & $\zprimaryjerkW$ & $\zsecondaryjerkW$ & $\zprobjerkW$ \\[1.5mm]
ab & $\zprimaryaccelab$ & $\zsecondaryaccelab$ & $\zprobaccelab$ & $\zprimaryjerkab$ & $\zsecondaryjerkab$ & $\zprobjerkab$ \\[1.5mm]
Y & $\zprimaryaccelY$ & $\zsecondaryaccelY$ & $\zprobaccelY$ & $\zprimaryjerkY$ & $\zsecondaryjerkY$ & $\zprobjerkY$ \\[1.5mm]
P & $\zprimaryaccelP$ & $\zsecondaryaccelP$ & $\zprobaccelP$ & $\zprimaryjerkP$ & $\zsecondaryjerkP$ & $\zprobjerkP$ \\[1.5mm]
M & $\zprimaryaccelM$ & $\zsecondaryaccelM$ & $\zprobaccelM$ & $\zprimaryjerkM$ & $\zsecondaryjerkM$ & $\zprobjerkM$ \\[1.5mm]
R & $\zprimaryaccelR$ & $\zsecondaryaccelR$ & $\zprobaccelR$ & $\zprimaryjerkR$ & $\zsecondaryjerkR$ & $\zprobjerkR$ \\[1.5mm]
O & $\zprimaryaccelO$ & $\zsecondaryaccelO$ & $\zprobaccelO$ & $\zprimaryjerkO$ & $\zsecondaryjerkO$ & $\zprobjerkO$ \\[1.5mm]
F & $\zprimaryaccelF$ & $\zsecondaryaccelF$ & $\zprobaccelF$ & $\zprimaryjerkF$ & $\zsecondaryjerkF$ & $\zprobjerkF$ \\[1.5mm]
ah & $\zprimaryaccelah$ & $\zsecondaryaccelah$ & $\zprobaccelah$ & $\zprimaryjerkah$ & $\zsecondaryjerkah$ & $\zprobjerkah$ \\[1.5mm]
af & $\zprimaryaccelaf$ & $\zsecondaryaccelaf$ & $\zprobaccelaf$ & $\zprimaryjerkaf$ & $\zsecondaryjerkaf$ & $\zprobjerkaf$ \\[1.5mm]
U & $\zprimaryaccelU$ & $\zsecondaryaccelU$ & $\zprobaccelU$ & $\zprimaryjerkU$ & $\zsecondaryjerkU$ & $\zprobjerkU$ \\[1.5mm]
L & $\zprimaryaccelL$ & $\zsecondaryaccelL$ & $\zprobaccelL$ & $\zprimaryjerkL$ & $\zsecondaryjerkL$ & $\zprobjerkL$ \\[1.5mm]
N & $\zprimaryaccelN$ & $\zsecondaryaccelN$ & $\zprobaccelN$ & $\zprimaryjerkN$ & $\zsecondaryjerkN$ & $\zprobjerkN$ \\[1.5mm]
ag & $\zprimaryaccelag$ & $\zsecondaryaccelag$ & $\zprobaccelag$ & $\zprimaryjerkag$ & $\zsecondaryjerkag$ & $\zprobjerkag$ \\[1.5mm]
aj & $\zprimaryaccelaj$ & $\zsecondaryaccelaj$ & $\zprobaccelaj$ & $\zprimaryjerkaj$ & $\zsecondaryjerkaj$ & $\zprobjerkaj$ \\[1.5mm]
V & $\zprimaryaccelV$ & $\zsecondaryaccelV$ & $\zprobaccelV$ & $\zprimaryjerkV$ & $\zsecondaryjerkV$ & $\zprobjerkV$ \\[1.5mm]
C & $\zprimaryaccelC$ & $\zsecondaryaccelC$ & $\zprobaccelC$ & $\zprimaryjerkC$ & $\zsecondaryjerkC$ & $\zprobjerkC$ \\[1.5mm]
G & $\zprimaryaccelG$ & $\zsecondaryaccelG$ & $\zprobaccelG$ & $\zprimaryjerkG$ & $\zsecondaryjerkG$ & $\zprobjerkG$ \\[1.5mm]
ai & $\zprimaryaccelai$ & $\zsecondaryaccelai$ & $\zprobaccelai$ & $\zprimaryjerkai$ & $\zsecondaryjerkai$ & $\zprobjerkai$ \\[1.5mm]
K & $\zprimaryaccelK$ & $\zsecondaryaccelK$ & $\zprobaccelK$ & $\zprimaryjerkK$ & $\zsecondaryjerkK$ & $\zprobjerkK$ \\[1.5mm]
aa & $\zprimaryaccelaa$ & $\zsecondaryaccelaa$ & $\zprobaccelaa$ & $\zprimaryjerkaa$ & $\zsecondaryjerkaa$ & $\zprobjerkaa$ \\[1.5mm]
H & $\zprimaryaccelH$ & $\zsecondaryaccelH$ & $\zprobaccelH$ & $\zprimaryjerkH$ & $\zsecondaryjerkH$ & $\zprobjerkH$ \\[1.5mm]
ad & $\zprimaryaccelad$ & $\zsecondaryaccelad$ & $\zprobaccelad$ & $\zprimaryjerkad$ & $\zsecondaryjerkad$ & $\zprobjerkad$ \\[1.5mm]
S & $\zprimaryaccelS$ & $\zsecondaryaccelS$ & $\zprobaccelS$ & $\zprimaryjerkS$ & $\zsecondaryjerkS$ & $\zprobjerkS$ \\[1.5mm]
ak & $\zprimaryaccelak$ & $\zsecondaryaccelak$ & $\zprobaccelak$ & $\zprimaryjerkak$ & $\zsecondaryjerkak$ & $\zprobjerkak$ \\[1.5mm]
ac & $\zprimaryaccelac$ & $\zsecondaryaccelac$ & $\zprobaccelac$ & $\zprimaryjerkac$ & $\zsecondaryjerkac$ & $\zprobjerkac$ \\[1.5mm]
Q & $\zprimaryaccelQ$ & $\zsecondaryaccelQ$ & $\zprobaccelQ$ & $\zprimaryjerkQ$ & $\zsecondaryjerkQ$ & $\zprobjerkQ$ \\[1.5mm]
E & $\zprimaryaccelE$ & $\zsecondaryaccelE$ & $\zprobaccelE$ & $\zprimaryjerkE$ & $\zsecondaryjerkE$ & $\zprobjerkE$ \\[1.5mm]
T & $\zprimaryaccelT$ & $\zsecondaryaccelT$ & $\zprobaccelT$ & $\zprimaryjerkT$ & $\zsecondaryjerkT$ & $\zprobjerkT$ \\[1.5mm]
X & $\zprimaryaccelX$ & $\zsecondaryaccelX$ & $\zprobaccelX$ & $\zprimaryjerkX$ & $\zsecondaryjerkX$ & $\zprobjerkX$ \\[1.5mm]
A & $\zprimaryaccelA$ & $\zsecondaryaccelA$ & $\zprobaccelA$ & $\zprimaryjerkA$ & $\zsecondaryjerkA$ & $\zprobjerkA$ \\[1.5mm]
D & $\zprimaryaccelD$ & $\zsecondaryaccelD$ & $\zprobaccelD$ & $\zprimaryjerkD$ & $\zsecondaryjerkD$ & $\zprobjerkD$ \\[1.5mm]
J & $\zprimaryaccelJ$ & $\zsecondaryaccelJ$ & $\zprobaccelJ$ & $\zprimaryjerkJ$ & $\zsecondaryjerkJ$ & $\zprobjerkJ$ \\[1.5mm]
\enddata
\label{table:results}
\tablecomments{ Final fits to the line of sight position for each pulsar. Pulsars are ordered by their projected distance from the centers of the cluster. We include the line of sight position for both the $l_1$ and $l_2$ position, as well as the relative probability between the two. Results are given for the acceleration-only fits as well as the acceleration plus jerk fitting. If the distributions for $l_1$ and $l_2$ are strongly mixed, or one solution is too small to detect, we only report statistics for the single observable peak in the posterior distribution for both the $l_1$ and $l_2$ solution.}
\end{deluxetable*}

\clearpage
\begin{appendix}
\section{A. Nearest Neighbor Accelerations}
\label{appendix:nn_accel}

Chandrasekhar (1943; hereafter C43) derives the {\it Holtzmark} probability distribution for acceleration due to an infinite distribution of point masses with mean number density $n$. This is accomplished by first computing the distribution for $N$ stars uniformly distributed within a radius $R$, and then letting $N\rightarrow \infty $ and $R \rightarrow \infty$  keeping the ratio $n=3N/4\pi R^3$ constant. A characteristic size of the acceleration, for a star of mass $m$, must be $a \sim Gmn^{2/3}$ by dimensional analysis. If the probability distribution for the mass of each star is $P(m)$, then the distribution for vector acceleration $\bvec{a}$ due to $N$ stars in radius $R$ is (C43 Equation 524 and 525)
\begin{eqnarray}
P(\bvec{a}) & = &  \Pi_{i=1}^N \int_{r_i<R} \frac{d^3x_i}{4\pi R^3/3} 
\int dm_i P(m_i) 
\delta^3 \left( \bvec{a} - \sum_{i=1}^N \frac{Gm_i \bvec{x}_i}{r_i^3} \right)
\label{eq:P_of_a}
\nonumber \\ & = & 
\int \frac{d^3s}{(2\pi)^3} e^{-i \bvec{s} \cdot \bvec{a}}
\left[ \int_{r<R} \frac{d^3x}{4\pi R^3/3} P(m) dm\ e^{iGm\bvec{x} \cdot \bvec{s} / r^3} \right]^N
\nonumber\\ & \rightarrow& 
\frac{1}{2\pi^2 a^3} \int_0^\infty dx x \sin(x) e^{(a_{\rm NN}x/a)^{3/2}}.
\label{eq:holtzmark}
\end{eqnarray}

In the second step, the $\delta$ function was expressed as a Fourier integral, and the same probability distribution is used for each of the $N$ stars. In the last step, the integrals over $d^3x$ as well as the ``acceleration wavenumber" $\bvec{s}$ have been computed analytically by C43, with the result given by his  Equation 549. The characteristic nearest-neighbor acceleration is
\begin{eqnarray}
a_{\rm NN} & = & 2\pi G \left( \frac{4}{15} \langle m^{3/2} \rangle n \right)^{2/3}\,,
\label{eq:aNN}
\end{eqnarray}
where $\langle m^{3/2} \rangle = \int dm P(m) m^{3/2}$. Equation \ref{eq:holtzmark} is the {\it Holtzmark} distribution for the acceleration due to an infinite sea of particles exerting a $1/r^2$ force. Note that $P(\bvec{a})$ is independent of the direction of the acceleration, as expected for an infinite uniform medium, and only depends on $a \equiv |\bvec{a}|$. The ratio of $a_{\rm NN}$ to the mean cluster acceleration, for a cluster of mass $M_{\rm cl} \sim Nm$ and radius $R_{\rm cl} \sim (N/n)^{1/3}$, and $N$ stars of mass $m$, can be estimated as
\begin{eqnarray}
\frac{a_{\rm NN}}{GM_{\rm cl}/R_{\rm cl}^2} & \sim & 10^{-2} \left( \frac{10^6}{N} \right)^{1/3}\,,
\end{eqnarray}
where $N$ is the number of stars in the cluster. Hence the mean cluster acceleration is expected to dominate over the nearest neighbor acceleration by a large margin.

Only accelerations $a \gg a_{\rm NN}$ may be comparable to that of the smooth cluster acceleration. C43 Equation 559 gives the expansion of our Equation \ref{eq:holtzmark} to be
\begin{eqnarray}
P(\bvec{a}) & \simeq & \frac{15}{32\pi a_{\rm NN}^3} \sqrt{ \frac{2}{\pi} } \left( \frac{a_{\rm NN}}{a} \right)^{9/2}.
\end{eqnarray}
This may be converted into a distribution for the line-of-sight acceleration $a_l$ by integrating over the perpendicular directions as
\begin{eqnarray}
P(a_l) & = & \int 2\pi a_\perp da_\perp P(\bvec{a})
= 2\pi \int_0^\infty da_\perp a_\perp\ \frac{15}{32\pi a_{\rm NN}^3} \sqrt{ \frac{2}{\pi} } \left( \frac{a^2_{\rm NN}}{a_\perp^2 + a_l^2} \right)^{9/4} =  \frac{3}{4} \sqrt{ \frac{2}{\pi} } \frac{1}{a_{\rm NN}} \left( \frac{ a_{\rm NN} }{\left| a_l \right|} \right)^{5/2}.
\end{eqnarray}
The integral of this distribution gives the cumulative probability, $P_{\rm c}(a_l)$, of finding a line-of-sight acceleration larger than a value $a_l$ to be
\begin{eqnarray}
P_{\rm c}(a_l) & = & \int_{a_l}^\infty da_l' P(a_l') = \frac{1}{\sqrt{2\pi} }\left( \frac{a_{\rm NN}}{ \left| a_l \right|} \right)^{3/2}.
\label{eqn:appendix_cum_prob}
\end{eqnarray}

This result shows that the likelihood that a near neighbor imparts an acceleration $a_l \gg a_{\rm NN}$ goes down steeply with $a_l$.

\clearpage
\section{B. Jerk Profile}
\label{appendix:jerks}

Appendix \ref{appendix:nn_accel} reviewed C43's derivation of the Holtzmark distribution for the acceleration due to an infinite medium of point masses. C43 goes on to derive a joint probability distribution for accelerations and jerks. However, in the discussion following our Equation \ref{eq:aNN}, it was shown that the mean potential of the cluster is likely much larger than the acceleration due to nearest neighbors, hence this joint distribution derived for an infinite medium is not useful in the present context. Instead the distribution for the jerks, regardless of the value of acceleration, is derived here.

A star of mass $m$ at position $\bvec{x}$ and with velocity $\bvec{v}$ produces a jerk
\begin{eqnarray}
\dot{\bvec{a}}  =  Gm \left( \frac{\bvec{v}}{r^3} - 3 \frac{\bvec{v} \cdot \bvec{x} \bvec{x}}{r^5} \right)
\end{eqnarray}
at the origin. The physical meaning of jerk is the rate of change gravitational acceleration due to the motion of the particle. The distribution of stars is taken to be uniform with mean number density $n$, and the velocities are given by a Maxwell Boltzmann distribution 
\begin{eqnarray}
P(\bvec{v}) d^3v & = & \frac{d^3v}{(2\pi \sigma^2)^{3/2}} e^{-v^2/2\sigma^2},
\end{eqnarray}
where $\sigma$ is the velocity dispersion. The jerk can then be estimated as
\begin{eqnarray}
\dot{a} & \sim & \frac{Gmv}{r^3} \sim Gm\sigma n.
\end{eqnarray}
\end{appendix}

\end{document}